\documentclass[aps, prl, superscriptaddress, floatfix, longbibliography, preprint]{revtex4-2}

\usepackage{amsfonts, amssymb, amsmath}
\usepackage{graphicx}
\usepackage{float}
\usepackage[plain]{fancyref}
\usepackage{placeins}
\usepackage{hyperref}
\usepackage{color,soul}
\usepackage{siunitx}
\usepackage{natbib}
\usepackage[version=3]{mhchem}
\usepackage{physics}
\usepackage{pdfrender}
\usepackage{mathtools}
\usepackage{cases}
\usepackage{verbatim}
\definecolor{myorange}{cmyk}{0.0149, 0.7523, 0.7816, 0.003}
\definecolor{mypurple}{cmyk}{0.6638, 0.9652, 0, 0}

\DeclareSIUnit\angstrom{\text {Å}}
\newcommand{\VQDi}{\ensuremath{V_\mathrm{QD1}}}
\newcommand{\VQDii}{\ensuremath{V_\mathrm{QD2}}}
\newcommand{\VQDiii}{\ensuremath{V_\mathrm{QD3}}}
\newcommand{\QDi}{\ensuremath{\mathrm{QD}_1}}
\newcommand{\QDii}{\ensuremath{\mathrm{QD}_2}}
\newcommand{\QDiii}{\ensuremath{\mathrm{QD}_3}}
\newcommand{\VHi}{\ensuremath{V_\mathrm{H1}}}
\newcommand{\VHii}{\ensuremath{V_\mathrm{H2}}}
\newcommand{\VL}{\ensuremath{V_\mathrm{L}}}
\newcommand{\VR}{\ensuremath{V_\mathrm{R}}}
\newcommand{\IL}{\ensuremath{I_\mathrm{L}}}
\newcommand{\IR}{\ensuremath{I_\mathrm{R}}}
\newcommand{\gL}{\ensuremath{g_\mathrm{L}}}

\newcommand{\Ibias}{\ensuremath{I_\mathrm{bias}}}

\newcommand{\Bx}{\ensuremath{B_\mathrm{x}}}

\newcommand{\nn}{\nonumber \\} 
\newcommand{\dg}{^{\dagger}}
\newcommand{\bpm}{\begin{pmatrix}}
\newcommand{\epm}{\end{pmatrix}}

\begin{document}

\title{Signatures of Majorana protection in a three-site Kitaev chain}

\author{Alberto~Bordin}
% \thanks{These authors contributed equally to this work.}
\author{Chun--Xiao~Liu}
\affiliation{QuTech and Kavli Institute of NanoScience, Delft University of Technology, 2600 GA Delft, The Netherlands}
\author{Tom~Dvir}
\affiliation{QuTech and Kavli Institute of NanoScience, Delft University of Technology, 2600 GA Delft, The Netherlands}
\affiliation{Quantum Machines, HaMasger St 35, Tel Aviv-Yafo, 6721407, Israel}
\author{Francesco~Zatelli}
\author{Sebastiaan~L.~D.~ten~Haaf}
\author{David~van~Driel}
\author{Guanzhong~Wang}
\author{Nick~van~Loo}
\author{Jan~Cornelis~Wolff}
\author{Thomas~van~Caekenberghe}
\author{Yining~Zhang}
\affiliation{QuTech and Kavli Institute of NanoScience, Delft University of Technology, 2600 GA Delft, The Netherlands}
\author{Ghada~Badawy}
\author{Sasa~Gazibegovic}
\author{Erik~P.A.M.~Bakkers}
\affiliation{Department of Applied Physics, Eindhoven University of Technology, 5600 MB Eindhoven, The Netherlands}
\author{Michael~Wimmer}
\author{Leo~P.~Kouwenhoven}
\author{Grzegorz~P.~Mazur}
\email{g.p.mazur@tudelft.nl}

\affiliation{QuTech and Kavli Institute of NanoScience, Delft University of Technology, 2600 GA Delft, The Netherlands}

% \normalsize{$^\dagger$ These authors contributed equally to this work.}

\date{\today}

\begin{abstract}
Majorana zero modes (MZMs) are non-Abelian excitations predicted to emerge at the edges of topological superconductors. One proposal for realizing a topological superconductor in one dimension involves a chain of spinless fermions, coupled through $p$-wave superconducting pairing and electron hopping. This concept is also known as the Kitaev chain. A minimal two-site Kitaev chain has recently been experimentally realized using quantum dots (QDs) coupled through a superconductor. In such a minimal chain, MZMs are quadratically protected against global perturbations of the QD electrochemical potentials. However, they are not protected from perturbations of the inter-QD couplings. In this work, we demonstrate that extending the chain to three sites offers greater protection than the two-site configuration. The enhanced protection is evidenced by the stability of the zero-energy modes, which is robust against variations in both the coupling amplitudes and the electrochemical potential variations in the constituent QDs. While our device offers all the desired control of the couplings it does not allow for superconducting phase control. Our experimental observations are in good agreement with numerical simulated conductances with phase averaging. Our work pioneers the development of longer Kitaev chains, a milestone towards topological protection in QD-based chains.
\end{abstract}

\maketitle

\section*{Introduction}
%Cite Kitaev first. 
The pursuit of topological superconductivity is driven by its potential for decoherence-free quantum computing~\cite{kitaev2001unpaired, Nayak2008RMP}. The primary strategy for realizing a topological superconductor has involved a method based on the Rashba nanowire model~\cite{lutchyn2010majorana,oreg2010helical}. 
This top-down approach relies on combining conventional $s$-wave superconductors with semiconducting nanowires with large spin-orbit coupling. These are expected to transition into a topological phase at certain chemical potentials in the presence of a sufficiently strong magnetic field. 
Over a decade of experimentation suggests that disorder is a primary obstacle hindering progress~\cite{ahn2021estimating,sarma2023spectral}, persisting even in state-of-the-art devices \cite{vanLoo2023,banerjee2023signatures,aghaee2022inas}.
An alternative, bottom-up approach involves constructing a topological superconductor using metamaterials. This can be achieved with a chain of quantum dots (QDs), coupled by crossed Andreev reflection (CAR) and elastic co-tunneling (ECT)~\cite{Sau.2012, Leijnse.2012}.
\newpage
\noindent
Until recently, controlling competing processes~\cite{schindele2012near} was not feasible.However, advances in device fabrication~\cite{Mazur.2022,moehle2021insbas} have not only enabled effective electrostatic control of these processes~\cite{liu2022tunable, bordin2023tunable,bordin2024crossed} but also led to the development of a new generation of extremely efficient Cooper pair splitters~\cite{Wang.2022, Wang.2022.2DEG} and minimal, two-site Kitaev chains~\cite{dvir2023realization,ten2023engineering, zatelli2023robust}. This rapid progress has sparked interest in using these systems for quantum information processing and fundamental physics experiments~\cite{spethmann2022coupled,Liu2023fusion,tsintzis2024majorana,pino2023minimal}.
\\[2ex]
In general, Kitaev chains allow local control over the electrochemical potentials of individual QDs, circumventing the issue of disorder. However, short Kitaev chains lack the topological protection offered by the top-down continuous long-wire proposal. Specifically, two-site chains are vulnerable to changes in the amplitudes of CAR and ECT, as theoretically predicted~\cite{Leijnse.2012} and experimentally demonstrated~\cite{dvir2023realization,zatelli2023robust}. Extending the chain to more sites can in principle offer protection against variation in CAR and ECT strengths, as well as further improve the stability of the chain against perturbation in the electrochemical potential. 

\section*{Coupling quantum dots}
\begin{figure}[ht!]
    \centering
    \includegraphics[width=\columnwidth]{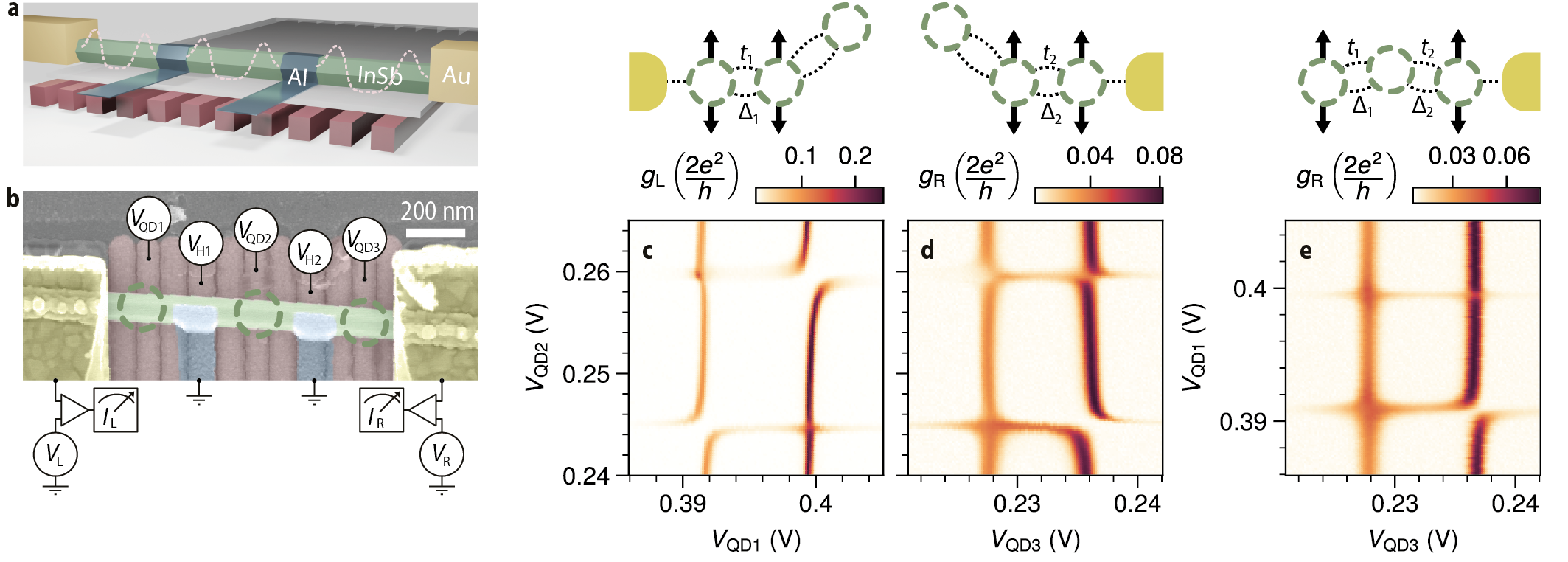}
    \caption{\textbf{a.} Illustrative diagram of the device. A semiconducting InSb nanowire (green) is placed on a keyboard of gates (pink), and contacted by two Al  (blue) and two Cr/Au (yellow) leads. The gates, separated from each other and from the nanowire by a thin dielectric, form a potential landscape defining three QDs, connected by two hybrid semiconducting-superconducting sections. \textbf{b.} False-colored scanning electron micrograph of the device. The superconductors are separately grounded through room temperature electronics, while the left and right normal probes are connected to corresponding voltage sources ($\VL$, $\VR$) and current meters ($\IL$, $\IR$). Differential conductances $\left( g_\mathrm{L} \equiv \frac{dI_\mathrm{L}}{dV_\mathrm{L}}, \; g_\mathrm{R} \equiv \frac{dI_\mathrm{R}}{dV_\mathrm{R}}\right)$ are measured with standard lockin techniques. Three green circles highlight the location of the QDs, controlled by the respective plunger gate voltages ($\VQDi$, $\VQDii$, $\VQDiii$) \textbf{c-e.} QD–QD charge stability diagrams. Zero bias conductance is measured across two charge degeneracy points for every pair of QDs. Avoided crossings indicate strong coupling between each pair. Crossings of the levels  signalize that couplings between the dots are equalized. In panel c, $\QDiii$ is kept off-resonance, in panel d, $\QDi$ is kept off-resonance, while, in panel e, $\QDii$ is set close to resonance, as the schematics above indicate.}
    \label{fig:1}
\end{figure}

\noindent
In order to engineer a three-site Kitaev chain Hamiltonian~\cite{kitaev2001unpaired, Sau.2012}
\begin{equation}
    H = \sum_{n=1}^3 \mu_n c_n^\dagger c_n + \sum_{n=1}^2 \left( t_n c_n^\dagger c_{n+1} + \Delta_n c_n^\dagger c_{n+1}^\dagger + h.c. \right),
    \label{eq:hamiltonian}
\end{equation}
where $c_n^\dagger$ and $c_n$ are the fermionic creation and annihilation operators, we need control over the onsite energies $\mu_n$, the hopping terms $t_n$ and the pairing terms $\Delta_n$. In our semiconducting nanowire device, shown in Fig~\ref{fig:1}a-b, three quantum dots are defined by an array of bottom gates, with $\VQDi$, $\VQDii$ and $\VQDiii$ controlling the electrochemical potentials $\mu_n$ of every QD. To lift the spin degeneracy, as prescribed by the Hamiltonian of Eq.~\ref{eq:hamiltonian}, we apply a magnetic field parallel to the nanowire axis ($\Bx = \SI{200}{mT}$). This leads to spin-polarization of the quantum dots. Tunneling spectroscopy of our semiconductor–superconductor hybrid segments is also performed at finite field and it is presented in the Supplementary Materials (Fig.~\ref{supp:ABS-spectroscopy}). In our previous work~\cite{bordin2024crossed}, we confirmed the presence of $t_{1,2}$ and $\Delta_{1,2}$ by detecting ECT and CAR across two hybrid segments with weakly coupled quantum dots. Here, we target strong couplings: $t_n, \Delta_n \gg k_B T$, where $k_B$ is the Boltzmann constant and $T$ the temperature. 
\newpage
\noindent
Indeed, the minimum value among $t_n$ and $\Delta_n$ determines the amplitude of the topological gap in a long Kitaev chain~\cite{Sau.2012,Fulga.2013}.To couple the QDs we rely on the Andreev Bound States (ABSs) populating the hybrids~\cite{liu2022tunable, bordin2023tunable, zatelli2023robust}. Measuring the zero-bias conductance on the left and the right of the device $\left( g_\mathrm{L} \equiv \frac{dI_\mathrm{L}}{dV_\mathrm{L}}, \; g_\mathrm{R} \equiv \frac{dI_\mathrm{R}}{dV_\mathrm{R}}\right)$,  we optimize the coupling site by site, as shown in the Supplementary Materials (Fig.~\ref{supp:s-shapes}), until we see the appearance of avoided crossings in the charge stability diagrams of Fig.~\ref{fig:1}c-e. 
Panels c and d show avoided crossings between $\QDi$ and $\QDii$ and between $\QDii$ and $\QDiii$ respectively. 
\newpage
\noindent
Remarkably, the coupling between neighboring QDs is strong enough to mediate interaction even between the outer QDs (panel e). We note that the coupling between $\QDi$ and $\QDiii$ is mediated by the middle QD as it is suppressed if $\QDii$ is moved off-resonance (see Fig.~\ref{supp:QD2-detuning}), indicating that the system Hamiltonian can be described by nearest-neighbor coupling terms giving rise to an effective next-nearest-neighbor coupling between the outer dots \cite{miles2023kitaev}.
\section*{Tuning 2-site Kitaev chains}

\begin{figure}[ht!]
    \centering
    \includegraphics[width=0.5\columnwidth]{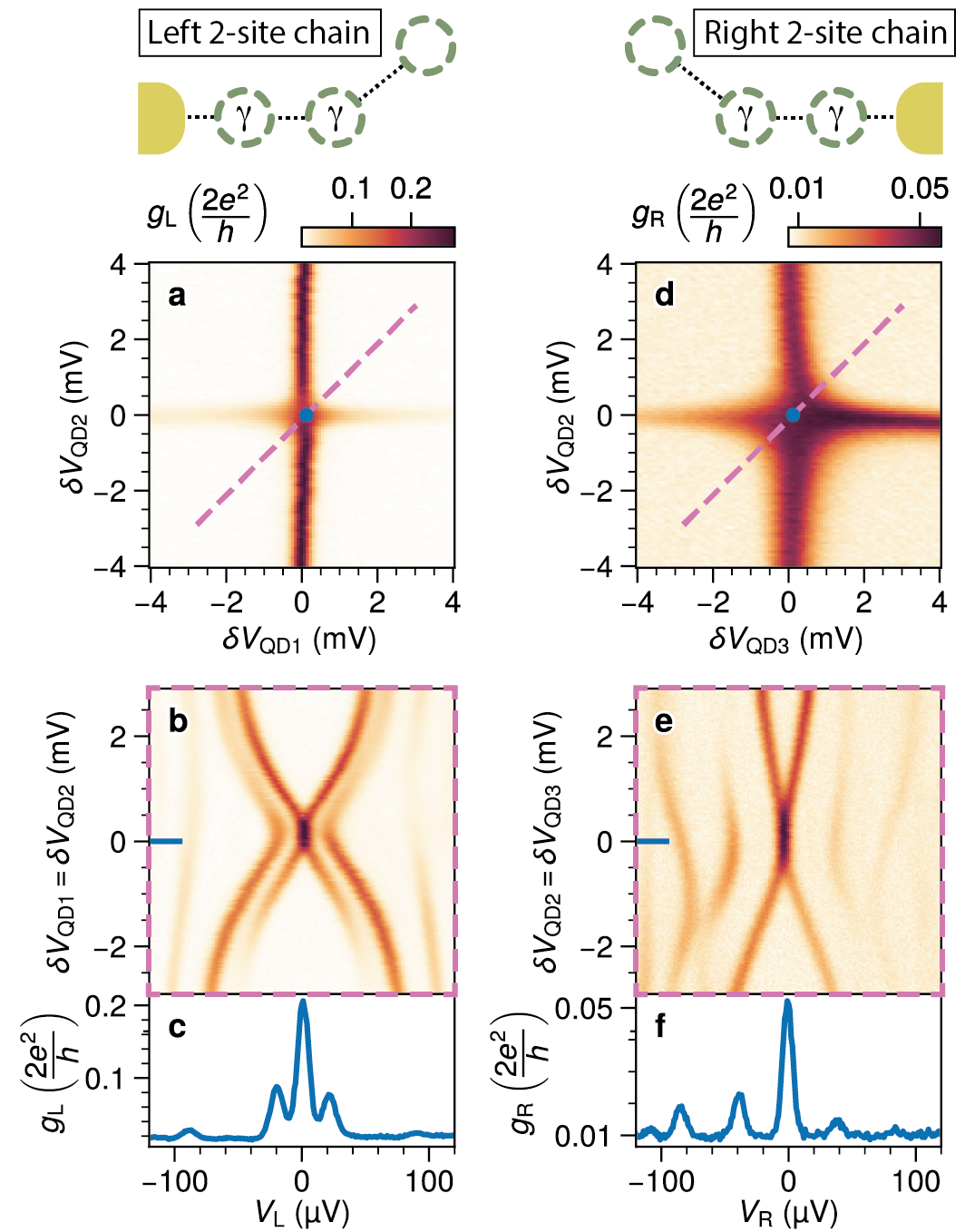}
    \caption{\textbf{Two-site Kitaev chains tuned on both ends of the device.} In the left column $\QDi$ and $\QDii$ are on resonance while $\QDiii$ is being kept off-resonance as depicted in the schematic. \textbf{a.} Charge stability diagram of $\QDi$ and $\QDii$ at a sweet spot where t$_{1}$=$\Delta_{1}$. \textbf{b.} Conductance spectroscopy as a function of simultaneous detuning $\delta$V of $\QDi$ and $\QDii$. \textbf{c.} Line-cut depicting spectrum at $\delta\VQDi=\delta\VQDii=0\,V$ illustrating a zero-bias peak and a gap of $\sim \SI{20}{\micro eV}$. Right column:  $\QDii$ and $\QDiii$ are kept on resonance, while $\QDi$ is kept off-resonance as depicted in the schematic. \textbf{d.} Charge stability diagram of $\QDii$ and $\QDiii$ at a sweet spot where t$_{2}$ = $\Delta_{2}$. \textbf{e.} Conductance spectroscopy as a function of simultaneous detuning $\delta V$ of $\QDii$ and $\QDiii$. \textbf{f.} Line-cut depicting spectrum at $\delta\VQDii$ = $\delta\VQDiii$ = 0\,V, illustrating a zero-bias peak and a gap of $\sim \SI{40}{\micro eV}$}
    \label{fig:2}
\end{figure}

After demonstrating strong coupling between the quantum dots, the next goal is to demonstrate the tunability of the chain. Ideally, ECT and CAR amplitudes should be balanced pairwise, setting:
\begin{equation}
    \Biggl\{
    \begin{aligned}
        t_1 &= \Delta_1 \\
        t_2 &= \Delta_2
    \end{aligned}
    \Bigr.
    \label{eq:double-ss}
\end{equation}
We begin by illustrating in Fig.~\ref{fig:2} how each condition of Eq.~\ref{eq:double-ss} can be individually met, with the constraint of keeping constant voltages on the gates forming $\QDii$. 
In the measurements of the left column of Fig.~\ref{fig:2}, $\QDiii$ is kept off-resonance, such that the low-energy behavior of the chain is effectively two sites. When $t_1 = \Delta_1$, we observe level crossing instead of repulsion in Fig.~\ref{fig:2}a~\cite{dvir2023realization}. The spectrum at the center, shown in Fig.~\ref{fig:2}c, shows a zero-bias conductance peak corresponding to a Poor Man's Majorana mode (PMM)~\cite{Leijnse.2012}, with excitation gap being $2t_1 = 2\Delta_1 \approx \SI{20}{\micro eV}$.  As pointed out in~\cite{Leijnse.2012}, if $\mu_1$ and $\mu_2$ are detuned from 0, then the PMMs split quadratically from zero energy, as shown in Fig~\ref{fig:2}b. Similarly, the right column of Fig.~\ref{fig:2} studies the case where $\QDi$ is kept off-resonance and the PMM pair appears on the right side of the device when $t_2 = \Delta_2$. We note that the gap of the right PMM pair is $\approx \SI{40}{\micro eV}$, twice as much compared to the left PMM pair. This is achieved with a higher degree of hybridization between the ABSs of the right hybrid and the neighboring quantum dots~ (Fig.~\ref{supp:QD-spectroscopy}), resulting in higher coupling strengths and increased stability of the zero bias peak~\cite{zatelli2023robust}, as can be appreciated in Fig.~\ref{fig:2}e,f. Although it is possible to tune the amplitudes of $t_n$ and $\Delta_n$ to be all equal, we choose to focus on the scenario where they are equal only pairwise. This approach allows us to identify spectral features arising from different coupling values in the chain.

\section*{The three-site chain}
\begin{figure}[ht!]
    \centering
    \includegraphics[width=\columnwidth]{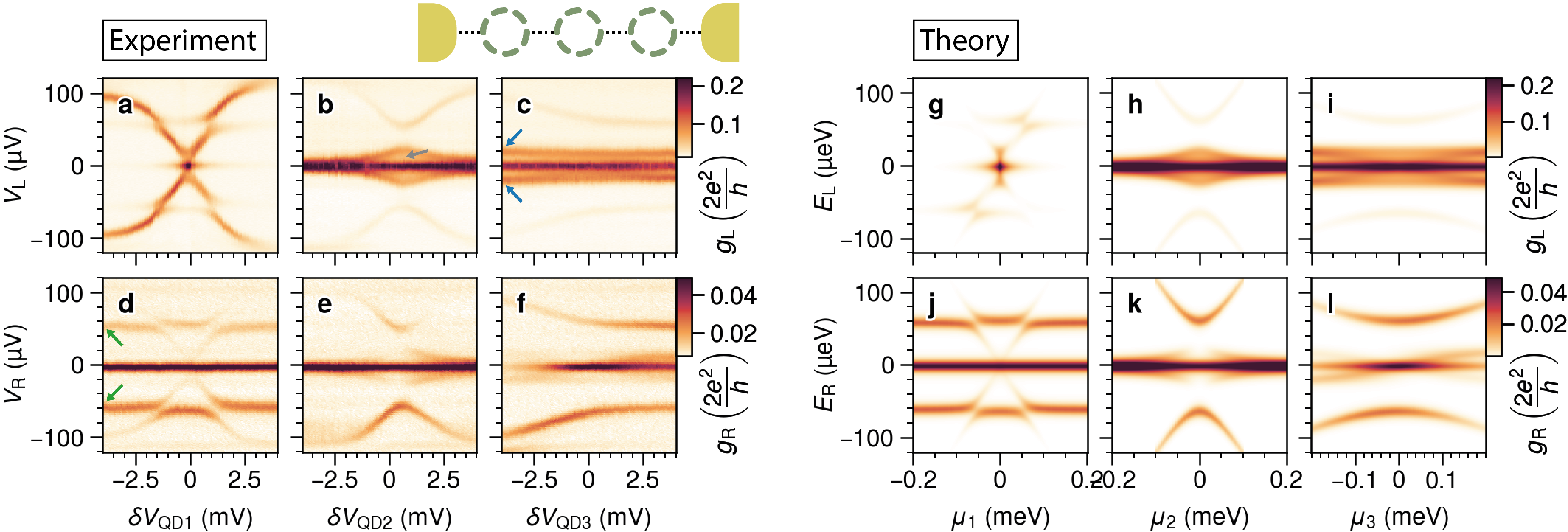}
    \caption{\textbf{a-c.} Conductance spectroscopy from the left lead, as a function of the detuning of individual quantum dots $\delta V_{QD_{n}}$ constituting the chain. By looking at the excited states when $\QDiii$ is off-resonance, we can estimate the left couplings to be $2t_1=2\Delta_1\approx\SI{24}{\micro eV}$ (blue arrows in panel c, $\delta \VQDiii = \SI{-4}{mV}$). \textbf{d-f.} Analogously, this section illustrates conductance spectroscopy from the right lead. When $\QDi$ is off-resonance we can estimate the left couplings to be $2t_2=2\Delta_2\approx\SI{60}{\micro eV}$ (green arrows in panel d). \textbf{g-l.} Each panel depicted here presents the results of numerical simulations corresponding to measurements presented in panels a to f.}
    \label{fig:3}
\end{figure}

Having satisfied the pairwise condition of Eq.~\ref{eq:double-ss}, we tune into the three-site Kitaev chain regime by setting all QDs on resonance. Fig.~\ref{fig:3} shows the spectrum of such a system, tunnel-probed from the left and the right (first and second row respectively), as a function of the detuning of every QD (first, second and third column).
\newpage
\noindent
The first observation is zero-bias conductance peaks manifesting on both ends of the device, remaining stable against the detuning of any constituent QD. Furthermore, spectroscopy from the left and the right reveal identical gate dispersions of the excited states, albeit with different intensities. Excited states originating from the left two sites are expected to couple more strongly to the left lead, while excited states originating from the right pair are expected to couple more strongly to the right one. Indeed, we identify excited states corresponding to $2t_1=2\Delta_1\approx\SI{24}{\micro eV}$, marked by blue arrows in Fig \ref{fig:3}c. Such states disperse only as a function of $\VQDi$ and $\VQDii$ and have higher $\gL$, signaling a higher local density of states. 
For the right side of the device, similar reasoning applies to the states marked by green arrows in Fig~\ref{fig:3}d, from which we estimate $2t_2=2\Delta_2\approx\SI{60}{\micro eV}$. 

Importantly, we observe that there is a finite conductance between the first excited state and the zero-bias peak (grey arrow in Fig~\ref{fig:3}b). While we have successfully equalized the amplitudes of the coupling parameters, another significant parameter to consider is the phase difference between them. In the Kitaev chain Hamiltonian (Eq.~\ref{eq:hamiltonian}), the terms $t_n$ and $\Delta_n$ are complex numbers, each with a distinct, non-trivial phase: $t_n = \abs{t_n}e^{i\phi_{n,t}}$ and $\Delta_n = \abs{\Delta_n}e^{i\phi_{n,\Delta}}$. In the context of a two-site Kitaev chain, the consideration of these phases is redundant as they can be absorbed into the quantum dot modes via a gauge transformation~\cite{Sau.2012}. The scenario changes however with a three-site Kitaev chain, where only three out of the four phases can be similarly absorbed, leaving one phase as an independent parameter. In our system, the phase difference originates from the superconducting leads, which then translates into the phase difference between $\Delta_1$ and $\Delta_2$, as explained in the Supplementary Material.
To understand the spectroscopic results presented in Fig.~\ref{fig:3}, we offer the following interpretation.Conceptually, the device's central part is a Josephson junction, that doesn't exhibit any measurable supercurrent when the device is tuned in a three-site chain configuration (see Fig.~\ref{supp:supercurrent}). As a result, the junction behaves ohmically and can support an infinitesimal voltage difference. According to the second Josephson relation~\cite{tinkham2004introduction}, finite voltage bias in Josephson junctions induces phase precession: $\frac{d\phi}{dt}=\frac{2eV}{\hbar}$. In our experiment, the voltage bias between the two superconducting leads cannot be set to zero with arbitrary precision, due to voltage divider effect, thermal fluctuations, finite equipment resolution and noise levels. %This is due to thermal voltages and drifts in the measurement electronics. 
We estimate the voltage difference to be on the order of $\delta V \sim \SI{1}{\micro V}$ (see Supplementary Material). The corresponding phase difference precesses with periods of $T_{\phi} =\frac{h}{2e \delta V} \sim 2$ ns. This is a very small time scale relative to the DC measurement time ($\sim \SI{1}{s}$). 
We thus assume that the spectra obtained for a three-site chain are uniformly averaged over possible phase differences.Fig.~\ref{fig:3}g-i shows the average simulated conductance of 25 phase selections uniformly distributed from 0 to $2\pi$. Within our interpretation, the zero-bias conductance peaks are still induced by Majorana zero modes. In particular, the Majoranas that should appear in a three-site Kitaev chain with zero phase difference would remain at zero energy regardless of the phase uncertainty. However, a major effect of the superconducting phase deviation is the change in the size of the energy gap, ranging from its maximum value $2t$ at 0-phase to 0 at $\pi$-phase. Our theoretical model reproduces the features observed in the experiment accurately, despite having only a few parameters. As opposed to a spinful model considering the ABSs in the hybrids explicitely~\cite{tsintzis2024majorana, liu2023enhancing}, the effective spinless model we are considering here only requires the fitting of the coupling to the leads $\Gamma_{L/R}$; all other model parameters are estimated from independent measurements (see Supplementary Material).
\newpage
\noindent
We note that these observations have been replicated also on another nanowire device with similar values of $t_n$ and $\Delta_n$, as presented in Fig.~\ref{supp:second-device}.

\section*{Increased protection}
\begin{figure}[ht!]
    \centering
    \includegraphics[width=0.66\columnwidth]{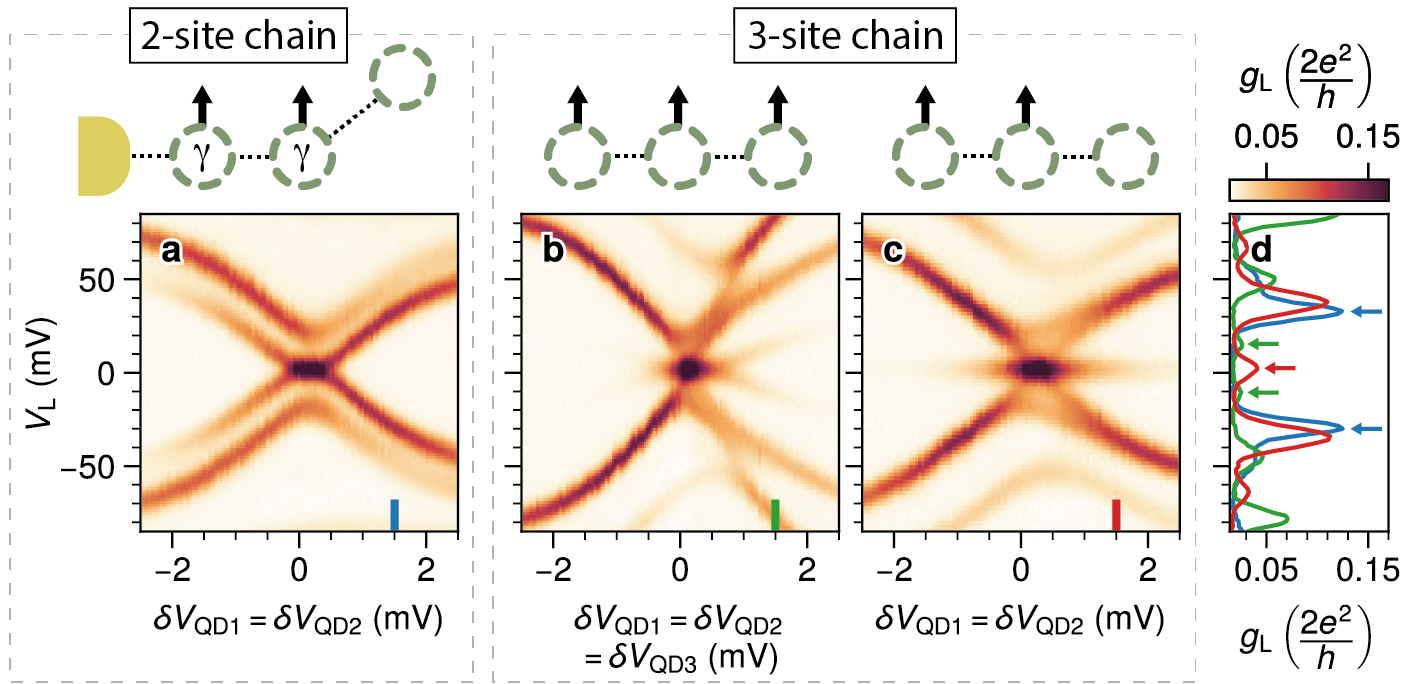}
    \caption{\textbf{Stability of a zero-energy state against electrochemical potential varation} \textbf{a.} Device in a 2-site chain configuration with $\QDiii$ out of resonance. Stability of a zero-energy state as a function of simultaneous detuning $\delta$V of $\QDi$ and $\QDii$. The data on panels b and c are taken in a 3-site chain configuration with $\QDiii$ brought back on resonance. \textbf{b.} Stability of a zero-energy state as a function of simultaneous detuning $\delta$V of QD$_{1,2,3}$, a visible splitting is observed once the dots are detuned by $\delta V = \SI{1.5}{mV}$. \textbf{c.} Conductance spectroscopy as a function of  $\delta$V of QD$_{1,2}$. In this case, the zero-energy state persists in the entire measurement range evidencing enhanced protection against local electro-chemical potential variation. \textbf{d.} Line-cuts of previous panels taken at $\delta V = \SI{1.5}{mV}$ for a 2-site chain (blue) and a three-site one with varying $\delta V_{1,2,3}$ (green) and $\delta V_{1,2}$ (red). Theory simulations are reported in Fig.~\ref{supp:protection-mu-theory}.}
    \label{fig:4}
\end{figure}

Fig.~\ref{fig:4} compares the robustness of two- and three-site chains against electrochemical potential variations. 
Notably, if all $\mu_n$ of a two-site chain are detuned, the zero modes should split quadratically, while, if all electrochemical potentials of a three-site chain are detuned, the zero modes should split cubically as we demonstrate in the Supplementary Material (Eq.~\ref{eq:cubic}). Panels a and b of Fig.~\ref{fig:4} show the comparison between these two scenarios (see also line-cuts in Fig.~\ref{fig:4}d). Moreover, panel c shows the spectroscopy of a three-site chain where only the first two chemical potentials are detuned. In this case, the zero bias peak persists over the full range. 
\newpage
We stress that in all panels we measure the conductance from the same left probe and, apart from $\VQDiii$, without changing any other bottom gate value between the scans. In particular, the gate settings of Fig.~\ref{fig:4}a and Fig.~\ref{fig:4}c are identical apart from the value of $\VQDiii$ ($\delta\VQDiii = \SI{-5}{mV}$ in Fig.~\ref{fig:4}a and $\SI{0}{mV}$ in panel Fig.~\ref{fig:4}c).

\begin{figure}[ht!]
    \centering
    \includegraphics[width=0.5\columnwidth]{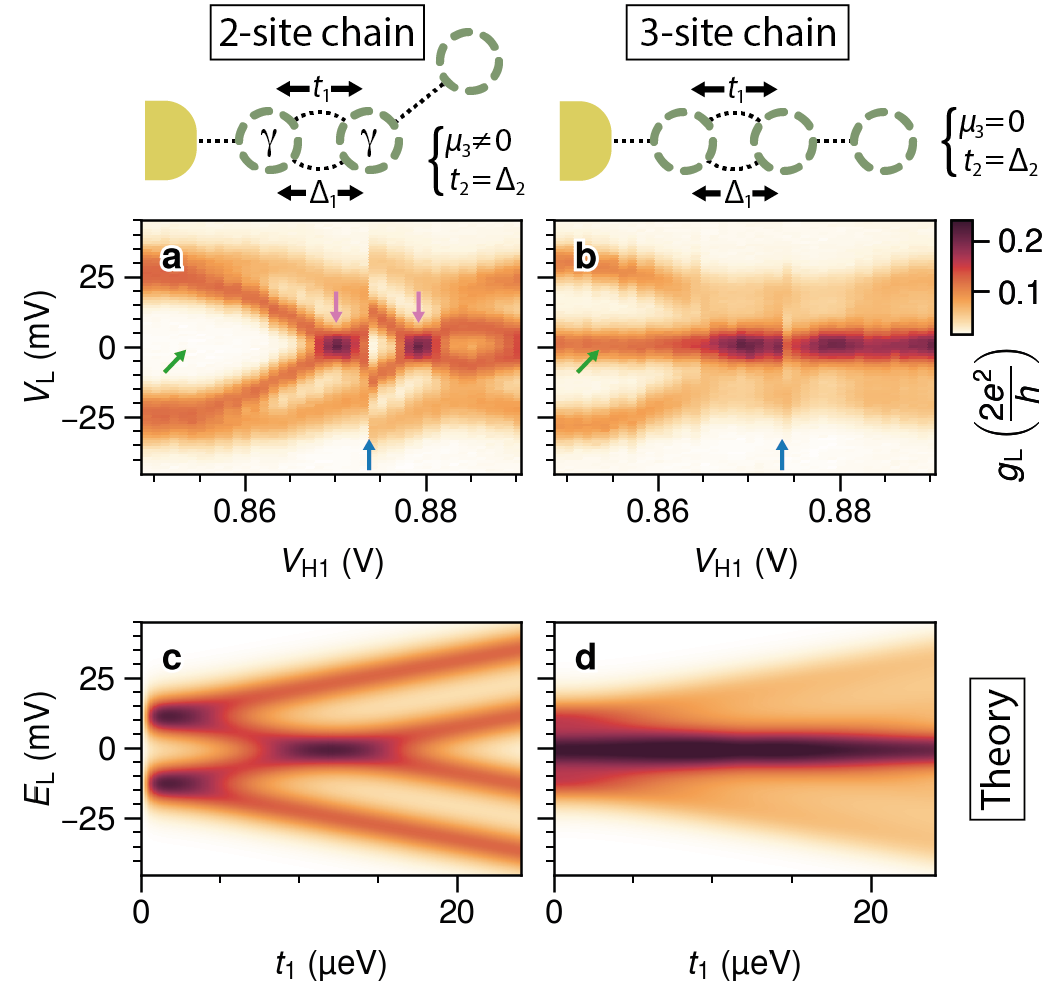}
    \caption{\textbf{Comparison of stability against variation of t and $\Delta$ in a two- and three-site Kitaev chain} \textbf{a.} Conductance spectroscopy of a two-site chain as a function of $\VHi$, which controls the magnitude of $t_{1}$ and $\Delta_{1}$. $\QDiii$ is kept off-resonance.
    \textbf{b.} $\QDiii$ is brought into resonance with the other two quantum dots in order to measure the spectrum of a three-site chain. Here, the zero-bias conductance peak persists over the entire $\VHi$ range. A blue arrow indicates a reproducible gate jump observed in this parameter region.
    \textbf{c, d.} Conductance simulation with $t_1$ varied from 0 to \SI{24}{\micro eV} for a 2-site chain (panel c) and a 3-site chain (panel d). $\Delta_1 = \SI{12}{\micro eV}$, $t_2 = \Delta_2 = \SI{30}{\micro eV}$. See also Figs.~\ref{supp:t-zoom}, \ref{supp:t2} and \ref{supp:protection-theory}.}
    \label{fig:5}
\end{figure}

It is noteworthy that in the pairwise sweet-spot condition of Eq.~\ref{eq:double-ss}, detuning two QDs is not enough to split the zero modes, but all three need to be detuned (Fig.~\ref{fig:4}b). Even in this case, the zero-bias peak splits at a lower rate compared to Fig.~\ref{fig:4}a. See Fig.~\ref{supp:bidiags} of the Supplementary Material for further details, where we show the persistence of the zero bias peak while detuning any pair of QDs.
Finally, Fig.~\ref{fig:5} compares the robustness of two- and three-site chains when leaving the pairwise sweet-spot condition of Eq.~\ref{eq:double-ss}. As opposed to electrochemical potential detuning, two-site chains have no protection against tunnel coupling deviations: perturbing either $t$ or $\Delta$ results in a linear splitting of the zero modes~\cite{Leijnse.2012, dvir2023realization, zatelli2023robust}. Here we reproduce such a result in panel a of Fig.~\ref{fig:5}. When $\QDiii$ is off-resonance, the zero bias peak of the left two-site chain is split as soon as the $\VHi$ controlling the $t_1$ and $\Delta_1$ ratio is detuned from the sweet-spot value (pink arrow). However, if we repeat the same measurement for the three-site Kitaev chain after bringing $\QDiii$ back on resonance, the zero-bias conductance peak persists over the entire $\VHi$ range (Fig.~\ref{fig:5}b), indicating tolerance to tunnel coupling deviations.

We note that the $\VHi$ range of Fig.~\ref{fig:5} is large enough to pass through a gate jump (blue arrow), which we find reproducible across multiple scans. While gate jumps can greatly affect the spectrum of 2-site chains, we find that a 3-site one is robust against them. Since the zero-bias peak persists, the gate jump clearly visible in panel a becomes barely noticeable in Fig.~\ref{fig:5}b.

As a summary, Figures~\ref{fig:4} and \ref{fig:5} show the increased protection of three-site chain Majorana zero modes compared to two-site chain ones. In particular, zero-modes in three site chains are more resilient against perturbations in the couplings $t_{n}$ and $\Delta_{n}$  (Fig.~\ref{fig:5}), which is expected to be the main limiting factor of coherence of Poor Man's Majorana-based qubits. The coherence time of a qubit made of 2-site Kitaev chains was previously predicted to be $\sim \SI{10}{ns}$~\cite{zatelli2023robust}. Although the coherence time of the three-site Kitaev chain in this work is limited by $T_{\phi}\sim \SI{2}{ns}$ due to Landau-Zener~\cite{Knapp2016prx} transitions near gap closing, based on the parameters extracted from the current experiment, in the Supplementary Material we estimate a qubit coherence time for a three-site Kitaev chain at $\phi=0$ to be around $\sim \SI{1}{\micro s}$. 
\newpage
This two orders of magnitude improvement provides further motivation for developing devices with phase control. By increasing the number of sites, the protection of MZMs against perturbations of $\mu_n$, $t_n$ and $\Delta_n$ is expected to further increase~\cite{Sau.2012}. In particular, we estimate that a 5-site chain would be enough for a target qubit lifetime of $\sim \SI{1}{ms}$ (Fig.~\ref{supp:more-sites}). 
Here we stress that while extending the chain length always leads to a quantitative enhancement in protection, the transition from two- to three-site also introduces a qualitative difference. Two-site chains lack protection against detuning in $t$ and $\Delta$, but a three-site chain represents the smallest chain where no single-parameter perturbation, by itself, can couple the two edge Majorana modes, split their energies and thus lead to decoherence. Ultimately, such protection comes from the additional middle QD acting as the ``bulk" of the chain.

\section*{Conclusion}
In this study, we have created a strongly coupled chain of three quantum dots, coupled through CAR  and ECT. Our devices have demonstrated the capability to host two adjacent two-site Kitaev chains. Additionally, we illustrate that when the three dots are on resonance, the system exhibits the spectrum expected for a three-site Kitaev chain, averaged across all possible phase differences.  Our setup permits the investigation of the MZM stability to variations in the electrochemical potential, as well as influences from CAR and ECT. This achievement addresses a key limitation of two-site Kitaev chains, where the finite overlap of MZM wavefunctions is considered a primary decoherence mechanism. In conclusion, extending Kitaev chains improves stability against $\mu_n$, $t_n$ and $\Delta_n$, appreciated even without the phase control, the next step towards qubit experiments.

\bibliography{bib.bib}

\begin{thebibliography}{36}%
\makeatletter
\providecommand \@ifxundefined [1]{%
 \@ifx{#1\undefined}
}%
\providecommand \@ifnum [1]{%
 \ifnum #1\expandafter \@firstoftwo
 \else \expandafter \@secondoftwo
 \fi
}%
\providecommand \@ifx [1]{%
 \ifx #1\expandafter \@firstoftwo
 \else \expandafter \@secondoftwo
 \fi
}%
\providecommand \natexlab [1]{#1}%
\providecommand \enquote  [1]{``#1''}%
\providecommand \bibnamefont  [1]{#1}%
\providecommand \bibfnamefont [1]{#1}%
\providecommand \citenamefont [1]{#1}%
\providecommand \href@noop [0]{\@secondoftwo}%
\providecommand \href [0]{\begingroup \@sanitize@url \@href}%
\providecommand \@href[1]{\@@startlink{#1}\@@href}%
\providecommand \@@href[1]{\endgroup#1\@@endlink}%
\providecommand \@sanitize@url [0]{\catcode `\\12\catcode `\$12\catcode `\&12\catcode `\#12\catcode `\^12\catcode `\_12\catcode `\%12\relax}%
\providecommand \@@startlink[1]{}%
\providecommand \@@endlink[0]{}%
\providecommand \url  [0]{\begingroup\@sanitize@url \@url }%
\providecommand \@url [1]{\endgroup\@href {#1}{\urlprefix }}%
\providecommand \urlprefix  [0]{URL }%
\providecommand \Eprint [0]{\href }%
\providecommand \doibase [0]{https://doi.org/}%
\providecommand \selectlanguage [0]{\@gobble}%
\providecommand \bibinfo  [0]{\@secondoftwo}%
\providecommand \bibfield  [0]{\@secondoftwo}%
\providecommand \translation [1]{[#1]}%
\providecommand \BibitemOpen [0]{}%
\providecommand \bibitemStop [0]{}%
\providecommand \bibitemNoStop [0]{.\EOS\space}%
\providecommand \EOS [0]{\spacefactor3000\relax}%
\providecommand \BibitemShut  [1]{\csname bibitem#1\endcsname}%
\let\auto@bib@innerbib\@empty
%</preamble>
\bibitem [{\citenamefont {Kitaev}(2001)}]{kitaev2001unpaired}%
  \BibitemOpen
  \bibfield  {author} {\bibinfo {author} {\bibfnamefont {A.~Y.}\ \bibnamefont {Kitaev}},\ }\href {https://doi.org/10.1070/1063-7869/44/10S/S29} {\bibfield  {journal} {\bibinfo  {journal} {Physics-uspekhi}\ }\textbf {\bibinfo {volume} {44}},\ \bibinfo {pages} {131} (\bibinfo {year} {2001})}\BibitemShut {NoStop}%
\bibitem [{\citenamefont {Nayak}\ \emph {et~al.}(2008)\citenamefont {Nayak}, \citenamefont {Simon}, \citenamefont {Stern}, \citenamefont {Freedman},\ and\ \citenamefont {Das~Sarma}}]{Nayak2008RMP}%
  \BibitemOpen
  \bibfield  {author} {\bibinfo {author} {\bibfnamefont {C.}~\bibnamefont {Nayak}}, \bibinfo {author} {\bibfnamefont {S.~H.}\ \bibnamefont {Simon}}, \bibinfo {author} {\bibfnamefont {A.}~\bibnamefont {Stern}}, \bibinfo {author} {\bibfnamefont {M.}~\bibnamefont {Freedman}},\ and\ \bibinfo {author} {\bibfnamefont {S.}~\bibnamefont {Das~Sarma}},\ }\href {https://doi.org/10.1103/RevModPhys.80.1083} {\bibfield  {journal} {\bibinfo  {journal} {Rev. Mod. Phys.}\ }\textbf {\bibinfo {volume} {80}},\ \bibinfo {pages} {1083} (\bibinfo {year} {2008})}\BibitemShut {NoStop}%
\bibitem [{\citenamefont {Lutchyn}\ \emph {et~al.}(2010)\citenamefont {Lutchyn}, \citenamefont {Sau},\ and\ \citenamefont {Sarma}}]{lutchyn2010majorana}%
  \BibitemOpen
  \bibfield  {author} {\bibinfo {author} {\bibfnamefont {R.~M.}\ \bibnamefont {Lutchyn}}, \bibinfo {author} {\bibfnamefont {J.~D.}\ \bibnamefont {Sau}},\ and\ \bibinfo {author} {\bibfnamefont {S.~D.}\ \bibnamefont {Sarma}},\ }\href@noop {} {\bibfield  {journal} {\bibinfo  {journal} {Physical review letters}\ }\textbf {\bibinfo {volume} {105}},\ \bibinfo {pages} {077001} (\bibinfo {year} {2010})}\BibitemShut {NoStop}%
\bibitem [{\citenamefont {Oreg}\ \emph {et~al.}(2010)\citenamefont {Oreg}, \citenamefont {Refael},\ and\ \citenamefont {Von~Oppen}}]{oreg2010helical}%
  \BibitemOpen
  \bibfield  {author} {\bibinfo {author} {\bibfnamefont {Y.}~\bibnamefont {Oreg}}, \bibinfo {author} {\bibfnamefont {G.}~\bibnamefont {Refael}},\ and\ \bibinfo {author} {\bibfnamefont {F.}~\bibnamefont {Von~Oppen}},\ }\href@noop {} {\bibfield  {journal} {\bibinfo  {journal} {Physical review letters}\ }\textbf {\bibinfo {volume} {105}},\ \bibinfo {pages} {177002} (\bibinfo {year} {2010})}\BibitemShut {NoStop}%
\bibitem [{\citenamefont {Ahn}\ \emph {et~al.}(2021)\citenamefont {Ahn}, \citenamefont {Pan}, \citenamefont {Woods}, \citenamefont {Stanescu},\ and\ \citenamefont {Sarma}}]{ahn2021estimating}%
  \BibitemOpen
  \bibfield  {author} {\bibinfo {author} {\bibfnamefont {S.}~\bibnamefont {Ahn}}, \bibinfo {author} {\bibfnamefont {H.}~\bibnamefont {Pan}}, \bibinfo {author} {\bibfnamefont {B.}~\bibnamefont {Woods}}, \bibinfo {author} {\bibfnamefont {T.~D.}\ \bibnamefont {Stanescu}},\ and\ \bibinfo {author} {\bibfnamefont {S.~D.}\ \bibnamefont {Sarma}},\ }\href@noop {} {\bibfield  {journal} {\bibinfo  {journal} {Physical Review Materials}\ }\textbf {\bibinfo {volume} {5}},\ \bibinfo {pages} {124602} (\bibinfo {year} {2021})}\BibitemShut {NoStop}%
\bibitem [{\citenamefont {Sarma}\ \emph {et~al.}(2023)\citenamefont {Sarma}, \citenamefont {Sau},\ and\ \citenamefont {Stanescu}}]{sarma2023spectral}%
  \BibitemOpen
  \bibfield  {author} {\bibinfo {author} {\bibfnamefont {S.~D.}\ \bibnamefont {Sarma}}, \bibinfo {author} {\bibfnamefont {J.~D.}\ \bibnamefont {Sau}},\ and\ \bibinfo {author} {\bibfnamefont {T.~D.}\ \bibnamefont {Stanescu}},\ }\href@noop {} {\bibfield  {journal} {\bibinfo  {journal} {arXiv preprint arXiv:2305.07007}\ } (\bibinfo {year} {2023})}\BibitemShut {NoStop}%
\bibitem [{\citenamefont {van Loo}\ \emph {et~al.}(2023)\citenamefont {van Loo}, \citenamefont {Mazur}, \citenamefont {Dvir}, \citenamefont {Wang}, \citenamefont {Dekker}, \citenamefont {Wang}, \citenamefont {Lemang}, \citenamefont {Sfiligoj}, \citenamefont {Bordin}, \citenamefont {van Driel} \emph {et~al.}}]{vanLoo2023}%
  \BibitemOpen
  \bibfield  {author} {\bibinfo {author} {\bibfnamefont {N.}~\bibnamefont {van Loo}}, \bibinfo {author} {\bibfnamefont {G.~P.}\ \bibnamefont {Mazur}}, \bibinfo {author} {\bibfnamefont {T.}~\bibnamefont {Dvir}}, \bibinfo {author} {\bibfnamefont {G.}~\bibnamefont {Wang}}, \bibinfo {author} {\bibfnamefont {R.~C.}\ \bibnamefont {Dekker}}, \bibinfo {author} {\bibfnamefont {J.-Y.}\ \bibnamefont {Wang}}, \bibinfo {author} {\bibfnamefont {M.}~\bibnamefont {Lemang}}, \bibinfo {author} {\bibfnamefont {C.}~\bibnamefont {Sfiligoj}}, \bibinfo {author} {\bibfnamefont {A.}~\bibnamefont {Bordin}}, \bibinfo {author} {\bibfnamefont {D.}~\bibnamefont {van Driel}}, \emph {et~al.},\ }\href@noop {} {\bibfield  {journal} {\bibinfo  {journal} {Nature Communications}\ }\textbf {\bibinfo {volume} {14}},\ \bibinfo {pages} {3325} (\bibinfo {year} {2023})}\BibitemShut {NoStop}%
\bibitem [{\citenamefont {Banerjee}\ \emph {et~al.}(2023)\citenamefont {Banerjee}, \citenamefont {Lesser}, \citenamefont {Rahman}, \citenamefont {Wang}, \citenamefont {Li}, \citenamefont {Kringh{\o}j}, \citenamefont {Whiticar}, \citenamefont {Drachmann}, \citenamefont {Thomas}, \citenamefont {Wang} \emph {et~al.}}]{banerjee2023signatures}%
  \BibitemOpen
  \bibfield  {author} {\bibinfo {author} {\bibfnamefont {A.}~\bibnamefont {Banerjee}}, \bibinfo {author} {\bibfnamefont {O.}~\bibnamefont {Lesser}}, \bibinfo {author} {\bibfnamefont {M.}~\bibnamefont {Rahman}}, \bibinfo {author} {\bibfnamefont {H.-R.}\ \bibnamefont {Wang}}, \bibinfo {author} {\bibfnamefont {M.-R.}\ \bibnamefont {Li}}, \bibinfo {author} {\bibfnamefont {A.}~\bibnamefont {Kringh{\o}j}}, \bibinfo {author} {\bibfnamefont {A.}~\bibnamefont {Whiticar}}, \bibinfo {author} {\bibfnamefont {A.}~\bibnamefont {Drachmann}}, \bibinfo {author} {\bibfnamefont {C.}~\bibnamefont {Thomas}}, \bibinfo {author} {\bibfnamefont {T.}~\bibnamefont {Wang}}, \emph {et~al.},\ }\href@noop {} {\bibfield  {journal} {\bibinfo  {journal} {Physical Review B}\ }\textbf {\bibinfo {volume} {107}},\ \bibinfo {pages} {245304} (\bibinfo {year} {2023})}\BibitemShut {NoStop}%
\bibitem [{\citenamefont {Aghaee}\ \emph {et~al.}(2022)\citenamefont {Aghaee}, \citenamefont {Akkala}, \citenamefont {Alam}, \citenamefont {Ali}, \citenamefont {Ramirez}, \citenamefont {Andrzejczuk}, \citenamefont {Antipov}, \citenamefont {Astafev}, \citenamefont {Bauer}, \citenamefont {Becker} \emph {et~al.}}]{aghaee2022inas}%
  \BibitemOpen
  \bibfield  {author} {\bibinfo {author} {\bibfnamefont {M.}~\bibnamefont {Aghaee}}, \bibinfo {author} {\bibfnamefont {A.}~\bibnamefont {Akkala}}, \bibinfo {author} {\bibfnamefont {Z.}~\bibnamefont {Alam}}, \bibinfo {author} {\bibfnamefont {R.}~\bibnamefont {Ali}}, \bibinfo {author} {\bibfnamefont {A.~A.}\ \bibnamefont {Ramirez}}, \bibinfo {author} {\bibfnamefont {M.}~\bibnamefont {Andrzejczuk}}, \bibinfo {author} {\bibfnamefont {A.~E.}\ \bibnamefont {Antipov}}, \bibinfo {author} {\bibfnamefont {M.}~\bibnamefont {Astafev}}, \bibinfo {author} {\bibfnamefont {B.}~\bibnamefont {Bauer}}, \bibinfo {author} {\bibfnamefont {J.}~\bibnamefont {Becker}}, \emph {et~al.},\ }\href@noop {} {\bibfield  {journal} {\bibinfo  {journal} {arXiv preprint arXiv:2207.02472}\ } (\bibinfo {year} {2022})}\BibitemShut {NoStop}%
\bibitem [{\citenamefont {Sau}\ and\ \citenamefont {Sarma}(2012)}]{Sau.2012}%
  \BibitemOpen
  \bibfield  {author} {\bibinfo {author} {\bibfnamefont {J.~D.}\ \bibnamefont {Sau}}\ and\ \bibinfo {author} {\bibfnamefont {S.~D.}\ \bibnamefont {Sarma}},\ }\href {https://doi.org/10.1038/ncomms1966} {\bibfield  {journal} {\bibinfo  {journal} {Nature Communications}\ }\textbf {\bibinfo {volume} {3}},\ \bibinfo {pages} {964} (\bibinfo {year} {2012})},\ \Eprint {https://arxiv.org/abs/1111.6600} {1111.6600} \BibitemShut {NoStop}%
\bibitem [{\citenamefont {Leijnse}\ and\ \citenamefont {Flensberg}(2012)}]{Leijnse.2012}%
  \BibitemOpen
  \bibfield  {author} {\bibinfo {author} {\bibfnamefont {M.}~\bibnamefont {Leijnse}}\ and\ \bibinfo {author} {\bibfnamefont {K.}~\bibnamefont {Flensberg}},\ }\href {https://doi.org/10.1103/physrevb.86.134528} {\bibfield  {journal} {\bibinfo  {journal} {Physical Review B}\ }\textbf {\bibinfo {volume} {86}},\ \bibinfo {pages} {134528} (\bibinfo {year} {2012})},\ \Eprint {https://arxiv.org/abs/1207.4299} {1207.4299} \BibitemShut {NoStop}%
\bibitem [{\citenamefont {Schindele}\ \emph {et~al.}(2012)\citenamefont {Schindele}, \citenamefont {Baumgartner},\ and\ \citenamefont {Sch{\"o}nenberger}}]{schindele2012near}%
  \BibitemOpen
  \bibfield  {author} {\bibinfo {author} {\bibfnamefont {J.}~\bibnamefont {Schindele}}, \bibinfo {author} {\bibfnamefont {A.}~\bibnamefont {Baumgartner}},\ and\ \bibinfo {author} {\bibfnamefont {C.}~\bibnamefont {Sch{\"o}nenberger}},\ }\href {https://doi.org/https://doi.org/10.1103/PhysRevLett.109.157002} {\bibfield  {journal} {\bibinfo  {journal} {Physical review letters}\ }\textbf {\bibinfo {volume} {109}},\ \bibinfo {pages} {157002} (\bibinfo {year} {2012})}\BibitemShut {NoStop}%
\bibitem [{\citenamefont {Mazur}\ \emph {et~al.}(2022)\citenamefont {Mazur}, \citenamefont {Loo}, \citenamefont {Wang}, \citenamefont {Dvir}, \citenamefont {Wang}, \citenamefont {Khindanov}, \citenamefont {Korneychuk}, \citenamefont {Borsoi}, \citenamefont {Dekker}, \citenamefont {Badawy}, \citenamefont {Vinke}, \citenamefont {Gazibegovic}, \citenamefont {Bakkers}, \citenamefont {Pérez}, \citenamefont {Heedt},\ and\ \citenamefont {Kouwenhoven}}]{Mazur.2022}%
  \BibitemOpen
  \bibfield  {author} {\bibinfo {author} {\bibfnamefont {G.~P.}\ \bibnamefont {Mazur}}, \bibinfo {author} {\bibfnamefont {N.~v.}\ \bibnamefont {Loo}}, \bibinfo {author} {\bibfnamefont {J.}~\bibnamefont {Wang}}, \bibinfo {author} {\bibfnamefont {T.}~\bibnamefont {Dvir}}, \bibinfo {author} {\bibfnamefont {G.}~\bibnamefont {Wang}}, \bibinfo {author} {\bibfnamefont {A.}~\bibnamefont {Khindanov}}, \bibinfo {author} {\bibfnamefont {S.}~\bibnamefont {Korneychuk}}, \bibinfo {author} {\bibfnamefont {F.}~\bibnamefont {Borsoi}}, \bibinfo {author} {\bibfnamefont {R.~C.}\ \bibnamefont {Dekker}}, \bibinfo {author} {\bibfnamefont {G.}~\bibnamefont {Badawy}}, \bibinfo {author} {\bibfnamefont {P.}~\bibnamefont {Vinke}}, \bibinfo {author} {\bibfnamefont {S.}~\bibnamefont {Gazibegovic}}, \bibinfo {author} {\bibfnamefont {E.~P. A.~M.}\ \bibnamefont {Bakkers}}, \bibinfo {author} {\bibfnamefont {M.~Q.}\ \bibnamefont {Pérez}}, \bibinfo {author} {\bibfnamefont {S.}~\bibnamefont {Heedt}},\ and\ \bibinfo {author} {\bibfnamefont
  {L.~P.}\ \bibnamefont {Kouwenhoven}},\ }\href {https://doi.org/10.1002/adma.202202034} {\bibfield  {journal} {\bibinfo  {journal} {Advanced Materials}\ ,\ \bibinfo {pages} {2202034}} (\bibinfo {year} {2022})},\ \Eprint {https://arxiv.org/abs/2202.10230} {2202.10230} \BibitemShut {NoStop}%
\bibitem [{\citenamefont {Moehle}\ \emph {et~al.}(2021)\citenamefont {Moehle}, \citenamefont {Ke}, \citenamefont {Wang}, \citenamefont {Thomas}, \citenamefont {Xiao}, \citenamefont {Karwal}, \citenamefont {Lodari}, \citenamefont {van~de Kerkhof}, \citenamefont {Termaat}, \citenamefont {Gardner} \emph {et~al.}}]{moehle2021insbas}%
  \BibitemOpen
  \bibfield  {author} {\bibinfo {author} {\bibfnamefont {C.~M.}\ \bibnamefont {Moehle}}, \bibinfo {author} {\bibfnamefont {C.~T.}\ \bibnamefont {Ke}}, \bibinfo {author} {\bibfnamefont {Q.}~\bibnamefont {Wang}}, \bibinfo {author} {\bibfnamefont {C.}~\bibnamefont {Thomas}}, \bibinfo {author} {\bibfnamefont {D.}~\bibnamefont {Xiao}}, \bibinfo {author} {\bibfnamefont {S.}~\bibnamefont {Karwal}}, \bibinfo {author} {\bibfnamefont {M.}~\bibnamefont {Lodari}}, \bibinfo {author} {\bibfnamefont {V.}~\bibnamefont {van~de Kerkhof}}, \bibinfo {author} {\bibfnamefont {R.}~\bibnamefont {Termaat}}, \bibinfo {author} {\bibfnamefont {G.~C.}\ \bibnamefont {Gardner}}, \emph {et~al.},\ }\href@noop {} {\bibfield  {journal} {\bibinfo  {journal} {Nano Letters}\ }\textbf {\bibinfo {volume} {21}},\ \bibinfo {pages} {9990} (\bibinfo {year} {2021})}\BibitemShut {NoStop}%
\bibitem [{\citenamefont {Liu}\ \emph {et~al.}(2022)\citenamefont {Liu}, \citenamefont {Wang}, \citenamefont {Dvir},\ and\ \citenamefont {Wimmer}}]{liu2022tunable}%
  \BibitemOpen
  \bibfield  {author} {\bibinfo {author} {\bibfnamefont {C.-X.}\ \bibnamefont {Liu}}, \bibinfo {author} {\bibfnamefont {G.}~\bibnamefont {Wang}}, \bibinfo {author} {\bibfnamefont {T.}~\bibnamefont {Dvir}},\ and\ \bibinfo {author} {\bibfnamefont {M.}~\bibnamefont {Wimmer}},\ }\href {https://doi.org/https://doi.org/10.1103/PhysRevLett.129.267701} {\bibfield  {journal} {\bibinfo  {journal} {Physical review letters}\ }\textbf {\bibinfo {volume} {129}},\ \bibinfo {pages} {267701} (\bibinfo {year} {2022})}\BibitemShut {NoStop}%
\bibitem [{\citenamefont {Bordin}\ \emph {et~al.}(2023)\citenamefont {Bordin}, \citenamefont {Wang}, \citenamefont {Liu}, \citenamefont {Ten~Haaf}, \citenamefont {Van~Loo}, \citenamefont {Mazur}, \citenamefont {Xu}, \citenamefont {Van~Driel}, \citenamefont {Zatelli}, \citenamefont {Gazibegovic} \emph {et~al.}}]{bordin2023tunable}%
  \BibitemOpen
  \bibfield  {author} {\bibinfo {author} {\bibfnamefont {A.}~\bibnamefont {Bordin}}, \bibinfo {author} {\bibfnamefont {G.}~\bibnamefont {Wang}}, \bibinfo {author} {\bibfnamefont {C.-X.}\ \bibnamefont {Liu}}, \bibinfo {author} {\bibfnamefont {S.~L.}\ \bibnamefont {Ten~Haaf}}, \bibinfo {author} {\bibfnamefont {N.}~\bibnamefont {Van~Loo}}, \bibinfo {author} {\bibfnamefont {G.~P.}\ \bibnamefont {Mazur}}, \bibinfo {author} {\bibfnamefont {D.}~\bibnamefont {Xu}}, \bibinfo {author} {\bibfnamefont {D.}~\bibnamefont {Van~Driel}}, \bibinfo {author} {\bibfnamefont {F.}~\bibnamefont {Zatelli}}, \bibinfo {author} {\bibfnamefont {S.}~\bibnamefont {Gazibegovic}}, \emph {et~al.},\ }\href@noop {} {\bibfield  {journal} {\bibinfo  {journal} {Physical Review X}\ }\textbf {\bibinfo {volume} {13}},\ \bibinfo {pages} {031031} (\bibinfo {year} {2023})}\BibitemShut {NoStop}%
\bibitem [{\citenamefont {Bordin}\ \emph {et~al.}(2024{\natexlab{a}})\citenamefont {Bordin}, \citenamefont {Li}, \citenamefont {Van~Driel}, \citenamefont {Wolff}, \citenamefont {Wang}, \citenamefont {Ten~Haaf}, \citenamefont {Wang}, \citenamefont {Van~Loo}, \citenamefont {Kouwenhoven},\ and\ \citenamefont {Dvir}}]{bordin2024crossed}%
  \BibitemOpen
  \bibfield  {author} {\bibinfo {author} {\bibfnamefont {A.}~\bibnamefont {Bordin}}, \bibinfo {author} {\bibfnamefont {X.}~\bibnamefont {Li}}, \bibinfo {author} {\bibfnamefont {D.}~\bibnamefont {Van~Driel}}, \bibinfo {author} {\bibfnamefont {J.~C.}\ \bibnamefont {Wolff}}, \bibinfo {author} {\bibfnamefont {Q.}~\bibnamefont {Wang}}, \bibinfo {author} {\bibfnamefont {S.~L.}\ \bibnamefont {Ten~Haaf}}, \bibinfo {author} {\bibfnamefont {G.}~\bibnamefont {Wang}}, \bibinfo {author} {\bibfnamefont {N.}~\bibnamefont {Van~Loo}}, \bibinfo {author} {\bibfnamefont {L.~P.}\ \bibnamefont {Kouwenhoven}},\ and\ \bibinfo {author} {\bibfnamefont {T.}~\bibnamefont {Dvir}},\ }\href@noop {} {\bibfield  {journal} {\bibinfo  {journal} {Physical Review Letters}\ }\textbf {\bibinfo {volume} {132}},\ \bibinfo {pages} {056602} (\bibinfo {year} {2024}{\natexlab{a}})}\BibitemShut {NoStop}%
\bibitem [{\citenamefont {Wang}\ \emph {et~al.}(2022)\citenamefont {Wang}, \citenamefont {Dvir}, \citenamefont {Mazur}, \citenamefont {Liu}, \citenamefont {van Loo}, \citenamefont {ten Haaf}, \citenamefont {Bordin}, \citenamefont {Gazibegovic}, \citenamefont {Badawy}, \citenamefont {Bakkers} \emph {et~al.}}]{Wang.2022}%
  \BibitemOpen
  \bibfield  {author} {\bibinfo {author} {\bibfnamefont {G.}~\bibnamefont {Wang}}, \bibinfo {author} {\bibfnamefont {T.}~\bibnamefont {Dvir}}, \bibinfo {author} {\bibfnamefont {G.~P.}\ \bibnamefont {Mazur}}, \bibinfo {author} {\bibfnamefont {C.-X.}\ \bibnamefont {Liu}}, \bibinfo {author} {\bibfnamefont {N.}~\bibnamefont {van Loo}}, \bibinfo {author} {\bibfnamefont {S.~L.}\ \bibnamefont {ten Haaf}}, \bibinfo {author} {\bibfnamefont {A.}~\bibnamefont {Bordin}}, \bibinfo {author} {\bibfnamefont {S.}~\bibnamefont {Gazibegovic}}, \bibinfo {author} {\bibfnamefont {G.}~\bibnamefont {Badawy}}, \bibinfo {author} {\bibfnamefont {E.~P.}\ \bibnamefont {Bakkers}}, \emph {et~al.},\ }\href {https://doi.org/https://doi.org/10.1038/s41586-022-05352-2} {\bibfield  {journal} {\bibinfo  {journal} {Nature}\ }\textbf {\bibinfo {volume} {612}},\ \bibinfo {pages} {448} (\bibinfo {year} {2022})}\BibitemShut {NoStop}%
\bibitem [{\citenamefont {Wang}\ \emph {et~al.}(2023)\citenamefont {Wang}, \citenamefont {Ten~Haaf}, \citenamefont {Kulesh}, \citenamefont {Xiao}, \citenamefont {Thomas}, \citenamefont {Manfra},\ and\ \citenamefont {Goswami}}]{Wang.2022.2DEG}%
  \BibitemOpen
  \bibfield  {author} {\bibinfo {author} {\bibfnamefont {Q.}~\bibnamefont {Wang}}, \bibinfo {author} {\bibfnamefont {S.~L.}\ \bibnamefont {Ten~Haaf}}, \bibinfo {author} {\bibfnamefont {I.}~\bibnamefont {Kulesh}}, \bibinfo {author} {\bibfnamefont {D.}~\bibnamefont {Xiao}}, \bibinfo {author} {\bibfnamefont {C.}~\bibnamefont {Thomas}}, \bibinfo {author} {\bibfnamefont {M.~J.}\ \bibnamefont {Manfra}},\ and\ \bibinfo {author} {\bibfnamefont {S.}~\bibnamefont {Goswami}},\ }\href@noop {} {\bibfield  {journal} {\bibinfo  {journal} {Nature Communications}\ }\textbf {\bibinfo {volume} {14}},\ \bibinfo {pages} {4876} (\bibinfo {year} {2023})}\BibitemShut {NoStop}%
\bibitem [{\citenamefont {Dvir}\ \emph {et~al.}(2023)\citenamefont {Dvir}, \citenamefont {Wang}, \citenamefont {Loo}, \citenamefont {Liu}, \citenamefont {Mazur}, \citenamefont {Bordin}, \citenamefont {Haaf}, \citenamefont {Wang}, \citenamefont {Driel}, \citenamefont {Zatelli}, \citenamefont {Li}, \citenamefont {Malinowski}, \citenamefont {Gazibegovic}, \citenamefont {Badawy}, \citenamefont {Bakkers}, \citenamefont {Wimmer},\ and\ \citenamefont {Kouwenhoven}}]{dvir2023realization}%
  \BibitemOpen
  \bibfield  {author} {\bibinfo {author} {\bibfnamefont {T.}~\bibnamefont {Dvir}}, \bibinfo {author} {\bibfnamefont {G.}~\bibnamefont {Wang}}, \bibinfo {author} {\bibfnamefont {N.~v.}\ \bibnamefont {Loo}}, \bibinfo {author} {\bibfnamefont {C.-X.}\ \bibnamefont {Liu}}, \bibinfo {author} {\bibfnamefont {G.~P.}\ \bibnamefont {Mazur}}, \bibinfo {author} {\bibfnamefont {A.}~\bibnamefont {Bordin}}, \bibinfo {author} {\bibfnamefont {S.~L. D.~t.}\ \bibnamefont {Haaf}}, \bibinfo {author} {\bibfnamefont {J.-Y.}\ \bibnamefont {Wang}}, \bibinfo {author} {\bibfnamefont {D.~v.}\ \bibnamefont {Driel}}, \bibinfo {author} {\bibfnamefont {F.}~\bibnamefont {Zatelli}}, \bibinfo {author} {\bibfnamefont {X.}~\bibnamefont {Li}}, \bibinfo {author} {\bibfnamefont {F.~K.}\ \bibnamefont {Malinowski}}, \bibinfo {author} {\bibfnamefont {S.}~\bibnamefont {Gazibegovic}}, \bibinfo {author} {\bibfnamefont {G.}~\bibnamefont {Badawy}}, \bibinfo {author} {\bibfnamefont {E.~P. A.~M.}\ \bibnamefont {Bakkers}}, \bibinfo {author} {\bibfnamefont
  {M.}~\bibnamefont {Wimmer}},\ and\ \bibinfo {author} {\bibfnamefont {L.~P.}\ \bibnamefont {Kouwenhoven}},\ }\href {https://doi.org/10.1038/s41586-022-05585-1} {\bibfield  {journal} {\bibinfo  {journal} {Nature}\ }\textbf {\bibinfo {volume} {614}},\ \bibinfo {pages} {445} (\bibinfo {year} {2023})}\BibitemShut {NoStop}%
\bibitem [{\citenamefont {ten Haaf}\ \emph {et~al.}(2023)\citenamefont {ten Haaf}, \citenamefont {Wang}, \citenamefont {Bozkurt}, \citenamefont {Liu}, \citenamefont {Kulesh}, \citenamefont {Kim}, \citenamefont {Xiao}, \citenamefont {Thomas}, \citenamefont {Manfra}, \citenamefont {Dvir} \emph {et~al.}}]{ten2023engineering}%
  \BibitemOpen
  \bibfield  {author} {\bibinfo {author} {\bibfnamefont {S.~L.}\ \bibnamefont {ten Haaf}}, \bibinfo {author} {\bibfnamefont {Q.}~\bibnamefont {Wang}}, \bibinfo {author} {\bibfnamefont {A.~M.}\ \bibnamefont {Bozkurt}}, \bibinfo {author} {\bibfnamefont {C.-X.}\ \bibnamefont {Liu}}, \bibinfo {author} {\bibfnamefont {I.}~\bibnamefont {Kulesh}}, \bibinfo {author} {\bibfnamefont {P.}~\bibnamefont {Kim}}, \bibinfo {author} {\bibfnamefont {D.}~\bibnamefont {Xiao}}, \bibinfo {author} {\bibfnamefont {C.}~\bibnamefont {Thomas}}, \bibinfo {author} {\bibfnamefont {M.~J.}\ \bibnamefont {Manfra}}, \bibinfo {author} {\bibfnamefont {T.}~\bibnamefont {Dvir}}, \emph {et~al.},\ }\href@noop {} {\bibfield  {journal} {\bibinfo  {journal} {arXiv preprint arXiv:2311.03208}\ } (\bibinfo {year} {2023})}\BibitemShut {NoStop}%
\bibitem [{\citenamefont {Zatelli}\ \emph {et~al.}(2023)\citenamefont {Zatelli}, \citenamefont {van Driel}, \citenamefont {Xu}, \citenamefont {Wang}, \citenamefont {Liu}, \citenamefont {Bordin}, \citenamefont {Roovers}, \citenamefont {Mazur}, \citenamefont {van Loo}, \citenamefont {Wolff} \emph {et~al.}}]{zatelli2023robust}%
  \BibitemOpen
  \bibfield  {author} {\bibinfo {author} {\bibfnamefont {F.}~\bibnamefont {Zatelli}}, \bibinfo {author} {\bibfnamefont {D.}~\bibnamefont {van Driel}}, \bibinfo {author} {\bibfnamefont {D.}~\bibnamefont {Xu}}, \bibinfo {author} {\bibfnamefont {G.}~\bibnamefont {Wang}}, \bibinfo {author} {\bibfnamefont {C.-X.}\ \bibnamefont {Liu}}, \bibinfo {author} {\bibfnamefont {A.}~\bibnamefont {Bordin}}, \bibinfo {author} {\bibfnamefont {B.}~\bibnamefont {Roovers}}, \bibinfo {author} {\bibfnamefont {G.~P.}\ \bibnamefont {Mazur}}, \bibinfo {author} {\bibfnamefont {N.}~\bibnamefont {van Loo}}, \bibinfo {author} {\bibfnamefont {J.~C.}\ \bibnamefont {Wolff}}, \emph {et~al.},\ }\href@noop {} {\bibfield  {journal} {\bibinfo  {journal} {arXiv preprint arXiv:2311.03193}\ } (\bibinfo {year} {2023})}\BibitemShut {NoStop}%
\bibitem [{\citenamefont {Spethmann}\ \emph {et~al.}(2022)\citenamefont {Spethmann}, \citenamefont {Zhang}, \citenamefont {Klinovaja},\ and\ \citenamefont {Loss}}]{spethmann2022coupled}%
  \BibitemOpen
  \bibfield  {author} {\bibinfo {author} {\bibfnamefont {M.}~\bibnamefont {Spethmann}}, \bibinfo {author} {\bibfnamefont {X.-P.}\ \bibnamefont {Zhang}}, \bibinfo {author} {\bibfnamefont {J.}~\bibnamefont {Klinovaja}},\ and\ \bibinfo {author} {\bibfnamefont {D.}~\bibnamefont {Loss}},\ }\href@noop {} {\bibfield  {journal} {\bibinfo  {journal} {Physical Review B}\ }\textbf {\bibinfo {volume} {106}},\ \bibinfo {pages} {115411} (\bibinfo {year} {2022})}\BibitemShut {NoStop}%
\bibitem [{\citenamefont {Liu}\ \emph {et~al.}(2023{\natexlab{a}})\citenamefont {Liu}, \citenamefont {Pan}, \citenamefont {Setiawan}, \citenamefont {Wimmer},\ and\ \citenamefont {Sau}}]{Liu2023fusion}%
  \BibitemOpen
  \bibfield  {author} {\bibinfo {author} {\bibfnamefont {C.-X.}\ \bibnamefont {Liu}}, \bibinfo {author} {\bibfnamefont {H.}~\bibnamefont {Pan}}, \bibinfo {author} {\bibfnamefont {F.}~\bibnamefont {Setiawan}}, \bibinfo {author} {\bibfnamefont {M.}~\bibnamefont {Wimmer}},\ and\ \bibinfo {author} {\bibfnamefont {J.~D.}\ \bibnamefont {Sau}},\ }\href {https://doi.org/10.1103/PhysRevB.108.085437} {\bibfield  {journal} {\bibinfo  {journal} {Phys. Rev. B}\ }\textbf {\bibinfo {volume} {108}},\ \bibinfo {pages} {085437} (\bibinfo {year} {2023}{\natexlab{a}})}\BibitemShut {NoStop}%
\bibitem [{\citenamefont {Tsintzis}\ \emph {et~al.}(2024)\citenamefont {Tsintzis}, \citenamefont {Souto}, \citenamefont {Flensberg}, \citenamefont {Danon},\ and\ \citenamefont {Leijnse}}]{tsintzis2024majorana}%
  \BibitemOpen
  \bibfield  {author} {\bibinfo {author} {\bibfnamefont {A.}~\bibnamefont {Tsintzis}}, \bibinfo {author} {\bibfnamefont {R.~S.}\ \bibnamefont {Souto}}, \bibinfo {author} {\bibfnamefont {K.}~\bibnamefont {Flensberg}}, \bibinfo {author} {\bibfnamefont {J.}~\bibnamefont {Danon}},\ and\ \bibinfo {author} {\bibfnamefont {M.}~\bibnamefont {Leijnse}},\ }\href@noop {} {\bibfield  {journal} {\bibinfo  {journal} {PRX Quantum}\ }\textbf {\bibinfo {volume} {5}},\ \bibinfo {pages} {010323} (\bibinfo {year} {2024})}\BibitemShut {NoStop}%
\bibitem [{\citenamefont {Pino}\ \emph {et~al.}(2024)\citenamefont {Pino}, \citenamefont {Souto},\ and\ \citenamefont {Aguado}}]{pino2023minimal}%
  \BibitemOpen
  \bibfield  {author} {\bibinfo {author} {\bibfnamefont {D.~M.}\ \bibnamefont {Pino}}, \bibinfo {author} {\bibfnamefont {R.~S.}\ \bibnamefont {Souto}},\ and\ \bibinfo {author} {\bibfnamefont {R.}~\bibnamefont {Aguado}},\ }\href {https://doi.org/10.1103/PhysRevB.109.075101} {\bibfield  {journal} {\bibinfo  {journal} {Phys. Rev. B}\ }\textbf {\bibinfo {volume} {109}},\ \bibinfo {pages} {075101} (\bibinfo {year} {2024})}\BibitemShut {NoStop}%
\bibitem [{\citenamefont {Fulga}\ \emph {et~al.}(2013)\citenamefont {Fulga}, \citenamefont {Haim}, \citenamefont {Akhmerov},\ and\ \citenamefont {Oreg}}]{Fulga.2013}%
  \BibitemOpen
  \bibfield  {author} {\bibinfo {author} {\bibfnamefont {I.~C.}\ \bibnamefont {Fulga}}, \bibinfo {author} {\bibfnamefont {A.}~\bibnamefont {Haim}}, \bibinfo {author} {\bibfnamefont {A.~R.}\ \bibnamefont {Akhmerov}},\ and\ \bibinfo {author} {\bibfnamefont {Y.}~\bibnamefont {Oreg}},\ }\href {https://doi.org/10.1088/1367-2630/15/4/045020} {\bibfield  {journal} {\bibinfo  {journal} {New journal of physics}\ }\textbf {\bibinfo {volume} {15}},\ \bibinfo {pages} {045020} (\bibinfo {year} {2013})}\BibitemShut {NoStop}%
\bibitem [{\citenamefont {Miles}\ \emph {et~al.}(2023)\citenamefont {Miles}, \citenamefont {van Driel}, \citenamefont {Wimmer},\ and\ \citenamefont {Liu}}]{miles2023kitaev}%
  \BibitemOpen
  \bibfield  {author} {\bibinfo {author} {\bibfnamefont {S.}~\bibnamefont {Miles}}, \bibinfo {author} {\bibfnamefont {D.}~\bibnamefont {van Driel}}, \bibinfo {author} {\bibfnamefont {M.}~\bibnamefont {Wimmer}},\ and\ \bibinfo {author} {\bibfnamefont {C.-X.}\ \bibnamefont {Liu}},\ }\href@noop {} {\bibfield  {journal} {\bibinfo  {journal} {arXiv preprint arXiv:2309.15777}\ } (\bibinfo {year} {2023})}\BibitemShut {NoStop}%
\bibitem [{\citenamefont {Tinkham}(2004)}]{tinkham2004introduction}%
  \BibitemOpen
  \bibfield  {author} {\bibinfo {author} {\bibfnamefont {M.}~\bibnamefont {Tinkham}},\ }\href@noop {} {\emph {\bibinfo {title} {{Introduction to Superconductivity}}}}\ (\bibinfo  {publisher} {Courier Corporation},\ \bibinfo {year} {2004})\BibitemShut {NoStop}%
\bibitem [{\citenamefont {Liu}\ \emph {et~al.}(2023{\natexlab{b}})\citenamefont {Liu}, \citenamefont {Bozkurt}, \citenamefont {Zatelli}, \citenamefont {ten Haaf}, \citenamefont {Dvir},\ and\ \citenamefont {Wimmer}}]{liu2023enhancing}%
  \BibitemOpen
  \bibfield  {author} {\bibinfo {author} {\bibfnamefont {C.-X.}\ \bibnamefont {Liu}}, \bibinfo {author} {\bibfnamefont {A.~M.}\ \bibnamefont {Bozkurt}}, \bibinfo {author} {\bibfnamefont {F.}~\bibnamefont {Zatelli}}, \bibinfo {author} {\bibfnamefont {S.~L.}\ \bibnamefont {ten Haaf}}, \bibinfo {author} {\bibfnamefont {T.}~\bibnamefont {Dvir}},\ and\ \bibinfo {author} {\bibfnamefont {M.}~\bibnamefont {Wimmer}},\ }\href@noop {} {\bibfield  {journal} {\bibinfo  {journal} {arXiv preprint arXiv:2310.09106}\ } (\bibinfo {year} {2023}{\natexlab{b}})}\BibitemShut {NoStop}%
\bibitem [{\citenamefont {Knapp}\ \emph {et~al.}(2016)\citenamefont {Knapp}, \citenamefont {Zaletel}, \citenamefont {Liu}, \citenamefont {Cheng}, \citenamefont {Bonderson},\ and\ \citenamefont {Nayak}}]{Knapp2016prx}%
  \BibitemOpen
  \bibfield  {author} {\bibinfo {author} {\bibfnamefont {C.}~\bibnamefont {Knapp}}, \bibinfo {author} {\bibfnamefont {M.}~\bibnamefont {Zaletel}}, \bibinfo {author} {\bibfnamefont {D.~E.}\ \bibnamefont {Liu}}, \bibinfo {author} {\bibfnamefont {M.}~\bibnamefont {Cheng}}, \bibinfo {author} {\bibfnamefont {P.}~\bibnamefont {Bonderson}},\ and\ \bibinfo {author} {\bibfnamefont {C.}~\bibnamefont {Nayak}},\ }\href {https://doi.org/10.1103/PhysRevX.6.041003} {\bibfield  {journal} {\bibinfo  {journal} {Phys. Rev. X}\ }\textbf {\bibinfo {volume} {6}},\ \bibinfo {pages} {041003} (\bibinfo {year} {2016})}\BibitemShut {NoStop}%
\bibitem [{\citenamefont {Scarlino}\ \emph {et~al.}(2022)\citenamefont {Scarlino}, \citenamefont {Ungerer}, \citenamefont {van Woerkom}, \citenamefont {Mancini}, \citenamefont {Stano}, \citenamefont {M\"uller}, \citenamefont {Landig}, \citenamefont {Koski}, \citenamefont {Reichl}, \citenamefont {Wegscheider}, \citenamefont {Ihn}, \citenamefont {Ensslin},\ and\ \citenamefont {Wallraff}}]{PhysRevX.12.031004}%
  \BibitemOpen
  \bibfield  {author} {\bibinfo {author} {\bibfnamefont {P.}~\bibnamefont {Scarlino}}, \bibinfo {author} {\bibfnamefont {J.~H.}\ \bibnamefont {Ungerer}}, \bibinfo {author} {\bibfnamefont {D.~J.}\ \bibnamefont {van Woerkom}}, \bibinfo {author} {\bibfnamefont {M.}~\bibnamefont {Mancini}}, \bibinfo {author} {\bibfnamefont {P.}~\bibnamefont {Stano}}, \bibinfo {author} {\bibfnamefont {C.}~\bibnamefont {M\"uller}}, \bibinfo {author} {\bibfnamefont {A.~J.}\ \bibnamefont {Landig}}, \bibinfo {author} {\bibfnamefont {J.~V.}\ \bibnamefont {Koski}}, \bibinfo {author} {\bibfnamefont {C.}~\bibnamefont {Reichl}}, \bibinfo {author} {\bibfnamefont {W.}~\bibnamefont {Wegscheider}}, \bibinfo {author} {\bibfnamefont {T.}~\bibnamefont {Ihn}}, \bibinfo {author} {\bibfnamefont {K.}~\bibnamefont {Ensslin}},\ and\ \bibinfo {author} {\bibfnamefont {A.}~\bibnamefont {Wallraff}},\ }\href {https://doi.org/10.1103/PhysRevX.12.031004} {\bibfield  {journal} {\bibinfo  {journal} {Phys. Rev. X}\ }\textbf {\bibinfo {volume} {12}},\ \bibinfo
  {pages} {031004} (\bibinfo {year} {2022})}\BibitemShut {NoStop}%
\bibitem [{\citenamefont {Heedt}\ \emph {et~al.}(2021)\citenamefont {Heedt}, \citenamefont {Quintero-Pérez}, \citenamefont {Borsoi}, \citenamefont {Fursina}, \citenamefont {Loo}, \citenamefont {Mazur}, \citenamefont {Nowak}, \citenamefont {Ammerlaan}, \citenamefont {Li}, \citenamefont {Korneychuk}, \citenamefont {Shen}, \citenamefont {Poll}, \citenamefont {Badawy}, \citenamefont {Gazibegovic}, \citenamefont {Jong}, \citenamefont {Aseev}, \citenamefont {Hoogdalem}, \citenamefont {Bakkers},\ and\ \citenamefont {Kouwenhoven}}]{Heedt.2021}%
  \BibitemOpen
  \bibfield  {author} {\bibinfo {author} {\bibfnamefont {S.}~\bibnamefont {Heedt}}, \bibinfo {author} {\bibfnamefont {M.}~\bibnamefont {Quintero-Pérez}}, \bibinfo {author} {\bibfnamefont {F.}~\bibnamefont {Borsoi}}, \bibinfo {author} {\bibfnamefont {A.}~\bibnamefont {Fursina}}, \bibinfo {author} {\bibfnamefont {N.~v.}\ \bibnamefont {Loo}}, \bibinfo {author} {\bibfnamefont {G.~P.}\ \bibnamefont {Mazur}}, \bibinfo {author} {\bibfnamefont {M.~P.}\ \bibnamefont {Nowak}}, \bibinfo {author} {\bibfnamefont {M.}~\bibnamefont {Ammerlaan}}, \bibinfo {author} {\bibfnamefont {K.}~\bibnamefont {Li}}, \bibinfo {author} {\bibfnamefont {S.}~\bibnamefont {Korneychuk}}, \bibinfo {author} {\bibfnamefont {J.}~\bibnamefont {Shen}}, \bibinfo {author} {\bibfnamefont {M.~A. Y. v.~d.}\ \bibnamefont {Poll}}, \bibinfo {author} {\bibfnamefont {G.}~\bibnamefont {Badawy}}, \bibinfo {author} {\bibfnamefont {S.}~\bibnamefont {Gazibegovic}}, \bibinfo {author} {\bibfnamefont {N.~d.}\ \bibnamefont {Jong}}, \bibinfo {author} {\bibfnamefont
  {P.}~\bibnamefont {Aseev}}, \bibinfo {author} {\bibfnamefont {K.~v.}\ \bibnamefont {Hoogdalem}}, \bibinfo {author} {\bibfnamefont {E.~P. A.~M.}\ \bibnamefont {Bakkers}},\ and\ \bibinfo {author} {\bibfnamefont {L.~P.}\ \bibnamefont {Kouwenhoven}},\ }\href {https://doi.org/10.1038/s41467-021-25100-w} {\bibfield  {journal} {\bibinfo  {journal} {Nature Communications}\ }\textbf {\bibinfo {volume} {12}},\ \bibinfo {pages} {4914} (\bibinfo {year} {2021})},\ \Eprint {https://arxiv.org/abs/2007.14383} {2007.14383} \BibitemShut {NoStop}%
\bibitem [{\citenamefont {Bordin}\ \emph {et~al.}(2024{\natexlab{b}})\citenamefont {Bordin}, \citenamefont {Li}, \citenamefont {Van~Driel}, \citenamefont {Wolff}, \citenamefont {Wang}, \citenamefont {Ten~Haaf}, \citenamefont {Wang}, \citenamefont {Van~Loo}, \citenamefont {Kouwenhoven},\ and\ \citenamefont {Dvir}}]{bordin2024crossed-supplemental-material}%
  \BibitemOpen
  \bibfield  {author} {\bibinfo {author} {\bibfnamefont {A.}~\bibnamefont {Bordin}}, \bibinfo {author} {\bibfnamefont {X.}~\bibnamefont {Li}}, \bibinfo {author} {\bibfnamefont {D.}~\bibnamefont {Van~Driel}}, \bibinfo {author} {\bibfnamefont {J.~C.}\ \bibnamefont {Wolff}}, \bibinfo {author} {\bibfnamefont {Q.}~\bibnamefont {Wang}}, \bibinfo {author} {\bibfnamefont {S.~L.}\ \bibnamefont {Ten~Haaf}}, \bibinfo {author} {\bibfnamefont {G.}~\bibnamefont {Wang}}, \bibinfo {author} {\bibfnamefont {N.}~\bibnamefont {Van~Loo}}, \bibinfo {author} {\bibfnamefont {L.~P.}\ \bibnamefont {Kouwenhoven}},\ and\ \bibinfo {author} {\bibfnamefont {T.}~\bibnamefont {Dvir}},\ }\href@noop {} {\bibinfo {title} {{Supplemental Material for ``Crossed Andreev Reflection and Elastic Cotunneling in Three Quantum Dots Coupled by Superconductors''}}},\ \bibinfo {howpublished} {\url{http://link.aps.org/supplemental/10.1103/PhysRevLett.132.056602}} (\bibinfo {year} {2024}{\natexlab{b}})\BibitemShut {NoStop}%
\bibitem [{\citenamefont {Bennebroek~Evertsz'}(2023)}]{florian2023}%
  \BibitemOpen
  \bibfield  {author} {\bibinfo {author} {\bibfnamefont {F.}~\bibnamefont {Bennebroek~Evertsz'}},\ }\href@noop {} {\bibinfo {title} {{Supercurrent Modulation by Andreev Bound States in a Quantum Dot Josephson Junction}}} (\bibinfo {year} {2023})\BibitemShut {NoStop}%
\bibitem [{\citenamefont {Martinez}\ \emph {et~al.}(2021)\citenamefont {Martinez}, \citenamefont {P{\"o}schl}, \citenamefont {Hansen}, \citenamefont {van~de Poll}, \citenamefont {Vaitiek{\.e}nas}, \citenamefont {Higginbotham},\ and\ \citenamefont {Casparis}}]{martinez2021measurement}%
  \BibitemOpen
  \bibfield  {author} {\bibinfo {author} {\bibfnamefont {E.~A.}\ \bibnamefont {Martinez}}, \bibinfo {author} {\bibfnamefont {A.}~\bibnamefont {P{\"o}schl}}, \bibinfo {author} {\bibfnamefont {E.~B.}\ \bibnamefont {Hansen}}, \bibinfo {author} {\bibfnamefont {M.~A.~Y.}\ \bibnamefont {van~de Poll}}, \bibinfo {author} {\bibfnamefont {S.}~\bibnamefont {Vaitiek{\.e}nas}}, \bibinfo {author} {\bibfnamefont {A.~P.}\ \bibnamefont {Higginbotham}},\ and\ \bibinfo {author} {\bibfnamefont {L.}~\bibnamefont {Casparis}},\ }\href@noop {} {\bibfield  {journal} {\bibinfo  {journal} {arXiv preprint arXiv:2104.02671}\ } (\bibinfo {year} {2021})}\BibitemShut {NoStop}%
\end{thebibliography}%

\section*{Acknowledgements}
This work has been supported by the Dutch Organization for Scientific Research (NWO) and Microsoft Corporation Station Q. We thank Raymond Schouten, Olaf Bennigshof and Jason Mensingh for technical support. We acknowledge Anton Akhmerov for suggesting the fast phase-precession interpretation and Rubén Seoane Souto, Srijit Goswami and Qingzhen Wang for fruitful discussions. 

\section*{Author contributions}

AB, JCW, DvD and GPM fabricated the device, AB, GPM, TD, FZ and TvC measured the devices and analyzed the data, with help from SLDtH, YW, GW and NvL. GB, SG and EPAMB performed nanowire synthesis. CXL and MW performed the numerical simulations. AB and TD initiated and designed early phase of the project. GPM and LPK supervised experiments presented in this manuscript. AB, GPM, CXL and LPK prepared the manuscript with input from all authors.

\section*{Data availability}
All raw data in the publication and the analysis code used to generate figures are available at 
\url{https://doi.org/10.5281/zenodo.10709983}. 

\clearpage
\pagebreak
\newpage

\begin{center}
  \textbf{\LARGE Supplementary Material}
\end{center}

\section{Theoretical model and simulation}
The Hamiltonian of a three-site Kitaev chain is
\begin{align}
H_{K3} &= \mu_1 n_1 + \mu_2 n_2+ \mu_3 n_3 
+ t_1 (c\dg_2 c_1 + c\dg_1 c_2) + t_2(c\dg_3 c_2 + c\dg_2 c_3)  \nn
&+ \Delta_1 (c\dg_2 c\dg_1 + c_1 c_2) + \Delta_2 ( e^{i\phi} c\dg_3 c\dg_2 +e^{-i\phi} c_2 c_3).
\label{eq:H_K3}
\end{align}
Here $c_i$ is the annihilation operator of the orbital in dot $i$, $n_i=c_i^\dagger c_i$ is the occupancy, $\mu_i$ is the orbital energy relative to the superconductor Fermi energy.
$t_{i}$ and $\Delta_{i}$ are the normal and superconducting tunnelings between dots $i$ and $i+1$, and $\phi$ is the phase difference between the two superconducting leads.
Physically, $t$'s and $\Delta$'s are the ECT and CAR amplitudes mediated by the subgap Andreev bound states in the hybrid segments.
In the Nambu basis, the above Hamiltonian can be written as
\begin{align}
&H = \frac{1}{2} \Psi\dg \cdot h_{BdG} \cdot \Psi, \nn
&\Psi = ( c_1, c_2, c_3, c\dg_1, c\dg_2, c\dg_3 )^T, \nn
&h_{BdG} = 
\bpm
\mu_1 & t_{1} & 0 & 0 & -\Delta_{1} & 0 \\
t_{1} & \mu_2 & t_{2} & \Delta_{1} & 0 & -\Delta_{2} e^{i\phi} \\
0 & t_{2} & \mu_3 & 0 & \Delta_{2} e^{i\phi} & 0 \\
0 & \Delta_{1} & 0 & -\mu_1 & -t_{1} & 0 \\
 -\Delta_{1} & 0 & \Delta_2 e^{-i\phi} & -t_{1} & -\mu_2 & -t_{2} \\
0 & -\Delta_{2} e^{-i\phi} & 0 & 0 & -t_{2} & -\mu_3
\epm.
\end{align}
When the system is coupled to normal leads, the scattering matrix describing the transmission and reflection amplitudes between modes in the leads can be expressed by the Weidenmuller formula 
\begin{align} 
S(\omega) = 1 - i W\dg \left( \omega  - h_{BdG} + \frac{i}{2} WW\dg \right)^{-1} W
\end{align}
where the tunnel matrix $W$ is defined as 
\begin{align}
W = \text{diag} \left(\sqrt{\Gamma_L}, 0, \sqrt{\Gamma_R}, 
-\sqrt{\Gamma_L}, 0, -\sqrt{\Gamma_R} \right),
\end{align}
with $\Gamma_{L/R}$ being the dot-lead coupling strength on the left and right ends respectively.
At zero temperature, the differential conductance is expressed as
\begin{align}
G^{(0)}_{ij}(\omega) \equiv dI_{i} / dV_{j} =  \delta_{ ij } - \abs{ S^{ee}_{  ij }(\omega) }^2 +  \abs{ S^{he}_{ ij }(\omega) }^2
\end{align}
in unit of $e^2/h$. 
Here $i,j=1,2,3$, and $\omega$ denotes the bias energy in the leads. 
The finite-temperature conductance is obtained by a convolution between the zero-temperature one and the derivative of the Fermi distribution
\begin{align}
G^{T}_{ij}(\omega) = \int^{+\infty}_{-\infty} dE \frac{G^{(0)}_{ij}(E)}{ 4k_BT \cosh^2[(E-\omega)/2k_BT] }.
\end{align}
In performing the numerical simulations, we choose the coupling strengths to be $t_{1}=\Delta_{1}=12~\mu eV$, $t_{2}=\Delta_{2}=30~\mu eV$ based on the positions of the excited states shown in Fig.~\ref{fig:3}.
The electron temperature in the normal leads, $T \sim \SI{35}{mK}$, corresponds to a broadening $k_\mathrm{B}T \sim \SI{3}{\micro eV}$. 
The strengths of the lead-dot couplings are chosen to be $\Gamma_L = \SI{2}{\micro eV}$, $\Gamma_R = \SI{0.4}{\micro eV}$, such that the conductance values obtained in the numerical simulations are close to those in the experimental measurements.
Moreover, to capture the effects of lever arms strength differences in the three dots, we choose $\delta \mu_1 = \delta \mu, \delta \mu_2 = \delta \mu, \delta \mu_3 = 0.3~\delta \mu$.
Crucially, we notice that in the particular experimental devices studied in this work, since the voltage bias between the two superconducting leads cannot be set to zero precisely, $\SI{0.1}{\mu V} \lesssim \delta V \lesssim \SI{1}{\mu V}$, the phase difference precesses with periods of $\SI{2}{ns} \lesssim T_{\phi} \sim \frac{h}{2e \delta V} \lesssim \SI{20}{ns}$.
On the other hand, the lifetime of an electron spent in a quantum dot is at the order of $\tau_e \sim \hbar/\Gamma \sim \SI{1}{ns}$.
This is the time scale of a single event of electron tunneling giving electric current, which would take a random value of phase difference $\phi$ since $\tau_e$ is smaller than or of similar order as the period of the phase winding $T_{\phi}$.
On the other hand, both $\tau_e$ and $T_{\phi}$ are a very small time scale relative to the DC current measurement time ($\sim \SI{1}{s}$).
%On the other hand, the voltage bias is relatively small compared with the excitation energy gap in the system, i.e., $\delta V \ll t, \Delta$, except for a small regime close to $\phi=\pi$ where the gap vanishes, thus the dynamical effect due to the phase evolution can be neglected.
Therefore, any particular data point collected in the conductance measurement is an average over $\sim 10^9$ tunneling events with different possible phases.
Theoretically, we capture this effect by performing a phase average on the differential conductance as follows
\begin{align}
    \langle G^{T}_{ij}(\omega)  \rangle_{\phi} \equiv \int^{2\pi}_0 \frac{d\phi}{2\pi} G^{T}_{ij}(\omega, \phi).
\end{align}

The numerically calculated conductances shown in the main text are obtained by averaging over 25 values of phases evenly distributed between $0$ and $2\pi$.

\subsection{Enhanced protection}
For the Majorana zero modes at the sweet spot of an $N$-site Kitaev chain  ($t_i=\Delta_i, \mu_i=0$), its energy deviation against onsite chemical potential fluctuation can be expressed as
\begin{align} 
\delta E_{K_N} \equiv E_{odd,gs} - E_{even,gs} =  \mu_N \prod^{N-1}_{i=1} \frac{ \mu_i}{2t_i},
\end{align}
where $t_i=\Delta_i$ are the strengths of the normal and superconducting couplings between sites $i$ and $i+1$.
In particular, for a two-site Kitaev chain, the energy deviation is
\begin{align}
\delta E_{K_2} = \frac{  \mu_1  \mu_2 }{2t} = C_{K_2} \cdot  \tilde{\mu}^2 ,
\end{align}
where $\mu_i=\mu$ and $\tilde{\mu} =\mu / (\mu eV)$ is dimensionless.
Note that for a two-site Kitaev chain, the protection is quadratic.
Here, the unit of $C_{K_2}$ is $\mu eV$, the physical meaning of which is the energy deviation when both dot orbitals are away from the fermi energy by 1$~\mu $eV.
Using the values obtained in the measurements, i.e., $t_1=15~\mu$eV and $t_2=30~\mu$eV, we estimate that $C_2=3.33\times 10^{-2}$ and $1.66\times 10^{-2}$ for the left and right pairs of PMM, respectively. 
For an extended three-site Kitaev chain, the energy deviation becomes
\begin{align}
\delta E_{K_3} = \frac{  \mu_1  \mu_2 \mu_3 }{4t_1t_2} =C_{K_3} \cdot  \tilde{\mu}^3,
\label{eq:cubic}
\end{align}
which is a cubic protection.
Now $C_{K_3}=5.55\times 10^{-4}$, which is smaller than $C_{K_2}$ by two orders of magnitudes, indicating a significantly enhanced degree of protection in the three-site Kitaev chain than its two-site version.
Moreover, we illustrate the physical meaning of ``exponential protection'' in scaling up a Kitaev chain. 
Without loss of generality, assuming homogeneity of the model parameters ($t_i=\Delta_i=t\equiv E_g/2$ with $E_g$ being the energy gap), we have 
\begin{align}
\delta E_{K_N} =  \mu_N \prod^{N-1}_{i=1} \frac{ \mu_i}{2t_i} = \frac{\mu^N}{E^{N-1}_g}
= E_g \left( \frac{\mu}{E_g} \right)^N = E_g \exp\{-N \log(E_g/\mu)\}.
\end{align}
Physically, it means that when the onsite energies of all the $N$ dots are detuned from zero by the same amount $\mu \ll E_g$, the energy splitting of the Majorana zero modes will decrease exponentially fast with the increasing number of sites at a rate of $\log(E_g/\mu)$.

\subsection{Derivation of the effective three-site Kitaev chain model}
In this section, we derive the effective Hamiltonian of a three-site Kitaev chain from a more microscopic level.
For the three quantum dots, in the presence of large Zeeman spin splitting and Coulomb interaction, we can approximate it to a single spin-polarized orbital as below:
\begin{align}
    H_{Di} = \mu_i (n_{i\uparrow} + n_{i\downarrow}) + 2 E_Zn_{i\uparrow}  + U_D n_{i\uparrow} n_{i\downarrow} \approx \mu_i n_{i\downarrow},
\end{align}
for $i=1,2,3$ and $\mu_i \approx 0$.
Here in the derivation, we focus on the spin-down orbitals in all dots, but the analysis holds for other spin polarizations as well.
On the other hand, the hybrid segment hosts subgap Andreev Bound States, of which the Hamiltonian is
\begin{align}
    H_{Aj} = \mu_j( n_{j\uparrow} + n_{j\downarrow} ) + |\Delta^{\text{ind}}_j|( 
e^{i\phi_j} c\dg_{j\uparrow}c\dg_{j\downarrow} + e^{-i\phi_j} c_{j\downarrow}c_{j\uparrow} )
\end{align}
where $j=L,R$, and $|\Delta^{\text{ind}}_j|$ is the magnitude of the induced gap, and $\phi_j$ is the superconducting phase.
The coupling between the dot and the hybrid is described by the following tunnel Hamiltonian
\begin{align}
    H_{tunn} =  \sum_{i=1,2} \sum_{\sigma=\uparrow,\downarrow} \left( wc\dg_{i\sigma } d_{i\sigma} + \sigma w_{so} c\dg_{i\overline{\sigma} } d_{i\sigma}
    + wd\dg_{i+1\sigma } c_{i\sigma} + \sigma w_{so} d\dg_{i+1\overline{\sigma} } c_{i\sigma} \right) + h.c.,
\end{align}
where $w$ and $w_{so}$ are the tunneling amplitudes for spin-conserving and spin-flipping processes.
As shown in Ref.~\cite{liu2022tunable}, in the tunneling regime, i.e., $w,w_{so}\ll |\Delta^{\text{ind}}|$, the Andreev bound states in the hybrid will mediate normal and superconducting couplings of quantum dots via elastic cotunneling and crossed Andreev reflection processes, giving
\begin{align}
    & t_1 =( w^2 -  w^2_{so}) \frac{u^2_L - v^2_L}{E_{AL}}, \nn
    & \Delta_1 = w w_{so} \frac{2u_L v_L}{E_{AL}}e^{i\phi_L}, \nn
    & t_2 =( w^2 -  w^2_{so}) \frac{u^2_R - v^2_R}{E_{AR}}, \nn
    & \Delta_2 = w w_{so} \frac{2u_R v_R}{E_{AR}}e^{i\phi_R}, 
    \label{eq:t_Delta_micro}
\end{align}
where $u_{j}, v_j$ are the BCS coherence factors of the ABS, and $E_{Aj} = \sqrt{\mu^2_j + (\Delta^{\text{ind}}_j)^2}$ is the excitation energy.
Thereby, by varying the chemical potential of the ABS, we can obtain the sweet spot by balancing the normal and superconducting coupling strengths ($|t_i| = |\Delta_i|$).
Furthermore, by performing a gauge transformation on the dot orbitals, we can remove the possible phases in three couplings in Eq.~\eqref{eq:t_Delta_micro}, and obtain
\begin{align}
    & t_1 \to |t_1|, \quad \Delta_1 \to |\Delta_1|, \nn
    & t_2 \to |t_2|, \quad \Delta_2 \to |\Delta_2|e^{i\phi}, \nn
    & \phi = \phi_R - \phi_L + \text{arg}( t_1 ) + \text{arg}( t_2 ).
\end{align}
That is, the effect of the phase difference between the two superconducting leads is now completely absorbed in a single parameter $\phi$. 
We therefore justify the use of the effective Hamiltonian in Eq.~\eqref{eq:H_K3} as the low-energy description of the dot-hybrid array.
We emphasize that in performing numerical simulations of dot energy detuning, as shown in Figs.~\ref{fig:3} and~\ref{fig:4}, the couplings between dots are just denoted by $t_j,\Delta_j$, while when considering the effect of voltage change in the hybrid segment for Fig.~\ref{supp:protection-theory}e-f, we reintroduce the effect of ABS using Eq.~\eqref{eq:t_Delta_micro} for the couplings in order to capture the $\mu_A$ dependence features.

\subsection{Estimation of dephasing rate for the Kitaev chain qubit}
In this subsection, we perform a numerical estimation of the dephasing time of different types of Kitaev chain qubits, similar to Ref.~\cite{zatelli2023robust} in spirit.
In particular, we consider four different types of Kitaev chain qubits: two-site Kitaev chain in the weak coupling regime, two-site Kitaev chain based on Yu-Shiba-Rusinov states in the strong coupling regime, and three-site Kitaev chain at $\phi=0$.
A Kitaev chain qubit consists of two subsystems of Kitaev chains, hosting Majoranas $\gamma_{A1}, \gamma_{A2}$ and $\gamma_{B1}, \gamma_{B2}$, respectively.
Without loss of generality, with total fermion parity being even, the two-qubit states are defined as $|0 \rangle = |e_A, e_B\rangle$ and $|1 \rangle = |o_A, o_B\rangle$.
As a result, the effective Hamiltonian of such a qubit is
\begin{align}
    H = \frac{i}{2}(E_A \gamma_{A1} \gamma_{A2} + E_B \gamma_{B1} \gamma_{B2} )=E\sigma_z,
\end{align}
where the Pauli operator $\sigma_z = i\gamma_{A1} \gamma_{A2}=i\gamma_{B1} \gamma_{B2}$, and $E = (E_A + E_B)/2$ with $E_{A/B} = E_{odd,A/B} - E_{even,A/B}$.
Here we assume that the two Kitaev chain subsystems have the same physical properties, and we estimate the dephasing rate based on a single Kitaev chain.
For a typical energy fluctuation of $\delta E$, the dephasing rate is $1/T^*_2 \sim \delta E/\hbar$.
%Here, the main source of energy fluctuations comes from the fluctuations in the gate voltages $V_i$ for both quantum dots and the hybrid segments.
Generally, charge noise is the main source of qubit decoherence, which is dominated by fluctuations of charge impurities in the environment.
However, as shown in Ref.~\cite{PhysRevX.12.031004}, the charge impurity fluctuations can be equivalently described by fluctuations in the gate voltages.
Theoretically, the voltage fluctuations enter the Kitaev-chain Hamiltonian as follows:
\begin{align}
& \delta \mu_i = \alpha_i \cdot \delta V_i, \nn
& \delta t_j = \frac{\partial t_j}{\partial V_{H_j}} \cdot \delta V_j, 
\end{align}
with $\alpha_i$ being the lever arm of the $i$-th quantum dot. 
In the second formula the derivative is extracted from a single pair of PMM (Fig.~\ref{fig:5}a).
We emphasize that the fluctuations of $t_j$ and $\Delta_j$ are correlated because both of them are induced by the ABS in the hybrid, which is controlled by a single electrostatic gate.
Here as a first-order approximation, we assume that the fluctuations are on $t_j$ while $\Delta_j$ remains constant.
We generate 5000 different disorder realizations of the set of gate voltages and calculate the averaged $\delta E$ which eventually gives the dephasing rate.
The voltage fluctuations obey Gaussian distribution with mean zero and standard deviation $\delta V \sim \SI{10}{\micro eV}$, as discussed in a similar experimental device~\cite{zatelli2023robust}.
We emphasize that for two-site Kitaev chains, there are three independent sources of fluctuations, two for the dots, and one for the hybrid segment.
For three-site Kitaev chains, there are five independent fluctuations, three for the dots and two for the hybrids.
Our analysis considers three distinct scenarios: dephasing due to dot energies only, hybrid coupling only, and both of them.
The device parameters used in our numerical simulations and the results of the estimations are summarized in Table~\ref{tab:dephasing_rate}.

\begin{table}
  \centering
  \begin{tabular}{|c|c|c|c|}
    \hline
    \textbf{Device parameters} & \textbf{QD-PMM}~\cite{dvir2023realization} & \textbf{YSR-PMM }~\cite{zatelli2023robust} & \textbf{Kitaev-3 ($\phi=0$)}  \\
    \hline
    $\alpha_{QD}~[e]$  & $0.3$ & $0.04$ & $0.04$  \\
    \hline
    $\partial t / \partial V_H~[e]$ & $5\times10^{-3}$ & $5\times10^{-3}$ & $5\times10^{-3}$  \\
    \hline
    $t,\Delta~[\mu \mathrm{eV}]$ & 12  & 40  & 10(left), 20(right)  \\ 
     \hline
     $1/T^*_2~[\mathrm{MHz}]$ ($\mu$ noises)  & $\sim 350$  & $\sim2$  & $\sim 0.06$  \\
      \hline
      $1/T^*_2~[\mathrm{MHz}]$ ($t$ noises)  & $\sim60$   & $\sim60$  & 0  \\
      \hline
      $1/T^*_2~[\mathrm{MHz}]$ (all noises) & $\sim 360$  & $\sim60$  & $\sim1.2$  \\
      \hline
  \end{tabular}
  \caption{Estimation of dephasing rate for different types of Kitaev chain qubits, assuming charge noise to be the only source of noises.}
  \label{tab:dephasing_rate}
\end{table}

\clearpage
\pagebreak
\newpage

\section{Nanofabrication and setup}

Our hybrid nanowire devices have been fabricated by means of the shadow-wall lithography technique thoroughly described in ref.~\cite{Heedt.2021}. Specific details are described in the \textit{Device structure} paragraph of ref.~\cite{bordin2024crossed} and its Supplemental Material~\cite{bordin2024crossed-supplemental-material}, which reports a detailed description of the dilution refrigerator setup as well.

\section{Tuning procedures}

\subsection{Strong coupling}

We report here the tuning protocol we follow to achieve strong coupling between all QD pairs.
First, we form QDs that are weekly coupled as in ref.~\cite{bordin2024crossed}. Weakly coupled QDs have high tunneling barriers and sharp Coulomb diamonds, since the broadening due to a finite lifetime is smaller than the broadening due to temperature.
Secondly, we start to couple the QDs more and more by progressivley lowering the tunneling barriers between them. Since, in our system, the coupling between QDs is mediated by ABSs~\cite{liu2022tunable, bordin2023tunable}, to optimize the barrier height we look at QD-ABS charge stability diagrams~\cite{zatelli2023robust}. To optimize, for instance, the right tunneling barrier of $\QDi$, we measure the zero-bias conductance $\gL$ as a function of $\VQDi$ and $\VHi$. As long as $\QDi$ resonances are not affected by $\VHi$, the tunneling barrier is too high. So we lower the tunneling barrier by increasing the corresponding bottom gate voltage and measure the $\QDi$-ABS charge stability diagram again. When the QD resonance lines start to bend as a function of $\VHi$, then $\QDi$ and the ABS start to hybridize, indicating the onset of strong coupling.
We repeat this procedure four times, once for every tunneling barrier in between the QDs, as Fig.~\ref{supp:s-shapes} shows. 
Finally, we check that QD-QD charge stability diagrams show avoided crossings as in Fig.~\ref{fig:1}, indicating a strong coupling between each pair of QDs.

We note that our device doesn't have a normal-metal probe directly connected to $\QDii$. Therefore, we start by tuning the middle QD while the outer ones are not yet formed. 
When there is a single tunneling barrier separating, for instance, the right hybrid and the right probe, it is possible to perform tunneling spectroscopy of the right hybrid as Fig.~\ref{supp:ABS-spectroscopy}b shows; and it is also possible to probe $\QDii$ as long as the right bias $\VR$ is kept below the ABS energies. A possible electron transport mechanism from the right probe to $\QDii$ is co-tunneling via the ABS, or even direct tunneling if the $\QDii$ is hybridized with the ABS~\cite{florian2023}. Regardless of the specific mechanism, $\QDii$ can be probed from the right normal-metal lead, as panels b and c of Fig.~\ref{supp:s-shapes} demonstrate.
After tuning the tunneling barriers of $\QDii$ with the procedure described above, we form $\QDi$ and $\QDiii$ and tune their inner barriers in the same way, as can be seen in panels a and d of Fig.~\ref{supp:s-shapes}. The outer tunneling barriers, i.e. the left barrier of $\QDi$ and the right barrier of $\QDiii$, are kept high to ensure a low coupling to the normal leads.

Finally, to avoid complications regarding supercurrent, we tune our device keeping the $\QDii$ tunneling barriers high enough so that we don't measure supercurrent between the two superconducting leads (Fig.~\ref{supp:supercurrent}).

\subsection{Poor Man's Majorana sweet-spots}

After achieving strong coupling between the QDs, the system needs to be tuned to the pairwise sweet-spot condition of Eq.~\ref{eq:double-ss}. The procedure is similar to what is presented in ref.~\cite{dvir2023realization}. The balance between CAR and ECT is found by looking at the direction of the avoided crossings in the QD-QD charge stability diagrams. We note that if the QDs are strongly coupled to the ABSs as in ref.~\cite{zatelli2023robust}, CAR and ECT are not well-defined anymore but need to be generalized to even-like and odd-like pairings. Here we stick to the CAR/ECT nomenclature for clarity and reference further readings for the generalized concepts~\cite{zatelli2023robust, liu2023enhancing}. 
An avoided crossing along the positive diagonal indicates CAR dominance and an avoided crossing along the negative diagonal indicates ECT dominance. We select a $\QDi$-$\QDii$ charge degeneracy point where it is possible to range from CAR dominance to ECT dominance by varying $\VHi$~\cite{bordin2023tunable}. Similarly, we select a $\QDii$-$\QDiii$ charge degeneracy point where it is possible to range from CAR dominance to ECT dominance by varying $\VHii$, with the added constraint that the $\QDii$ resonance must be the same for both choices. This is an important point: to be able to combine the tuning of the left and right QD pairs into a three-site chain, the gate settings of $\QDii$ must be exactly the same for both pairs. To achieve this, we tune the left pair and the right pair iteratively, converging to a pairwise sweet-spot condition that shares the gate settings of $\QDii$.
For this reason, Fig.~\ref{fig:2} and Fig.~\ref{fig:3} share the same settings for all 11 bottom gates, apart, obviously, from QD$_{1,2,3}$ depending on the panel. We note a discrepancy between the estimation of $t_2 = \Delta_2$, which is $\sim \SI{40}{\micro eV}$ for Fig.~\ref{fig:2} and $\sim \SI{60}{\micro eV}$ for Fig.~\ref{fig:3}. We attribute such discrepancy to a small charge jump for the right tunneling gate of $\QDii$ between the two measurements.

When CAR and ECT are balanced for both pairs, the charge stability diagrams show crossings instead of avoided crossings and the spectrum measured at the charge degeneracy points show zero-bias peaks (Fig.~\ref{fig:2}). Away from such sweet-spots, the zero-bias peaks are split, as Fig.~\ref{fig:5}a, Fig.~\ref{supp:t-zoom}a and Fig.~\ref{supp:t2}a show.

\subsection*{Calibration of the voltage difference between the superconducting leads}

The superconducting leads of our device are separately grounded via room-temperature electronics. This facilitates the tuning and characterization of $\QDii$ as shown in ref.~\cite{bordin2024crossed}. For a precise calibration of the voltage offset between the two superconducting leads, we tune the device to sustain a measurable supercurrent (see Fig.~\ref{supp:supercurrent}b for an example). With zero voltage drop across the device, even a small offset between the room-temperature grounds translates into a measurable current due to the two $\approx \SI{3}{k \Omega}$ resistances of the dilution refrigerator DC lines. Looking at the sign of such currents and adjusting accordingly, we can calibrate the voltage offset between the grounds down to less than $\SI{50}{nV}$. Finally, since fluctuations in the room temperature and $1/f$ noise of the electronic equipment can affect the ground calibration, we repeat this procedure on different days and assess how much the offset can drift over time. For the second device reported in Fig.~\ref{supp:second-device}, such offset was on the order of $\sim \SI{1}{\micro V}$. For the first device, concerning all other figures, such offset was always lower than $\SI{1}{\micro V}$ and typically closer to $\sim \SI{0.1}{\micro V}$. Lastly, we note that a finite voltage applied to the left or right normal-metal leads ($\VL$ or $\VR$) might lead to an effective voltage difference between the two superconducting leads due to the voltage divider effect~\cite{martinez2021measurement}; we calculate the impact of such effect on the voltage offset between the superconductors to be $\sim \SI{0.1}{\micro V}$ as well.

\clearpage
\pagebreak
\newpage

\renewcommand\thefigure{S\arabic{figure}}
\setcounter{figure}{0}

\begin{figure}[ht!]
    \centering
    \includegraphics[width=\columnwidth]{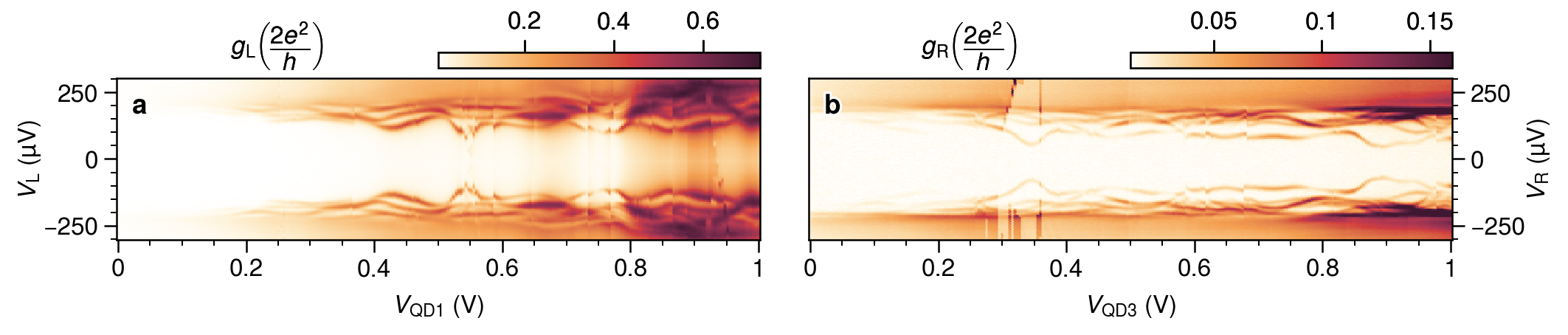}
    \caption{\textbf{ABS spectroscopy.} \textbf{a.} Spectroscopy of the left hybrid. \textbf{b.} Spectroscopy of the right hybrid. Both panels are measured at a fixed external magnetic field roughly parallel to the nanowire $\Bx = \SI{200}{mT}$ and exhibit ABSs populating the spectrum. We chose a magnetic field intensity that is large enough to polarize the dots ($\gtrsim \SI{100}{mT}$) but small enough for the ABSs not to close the gap ($\lesssim \SI{300}{mT}$).}
    \label{supp:ABS-spectroscopy}
\end{figure}

\begin{figure}[ht!]
    \centering
    \includegraphics[width=0.8\columnwidth]{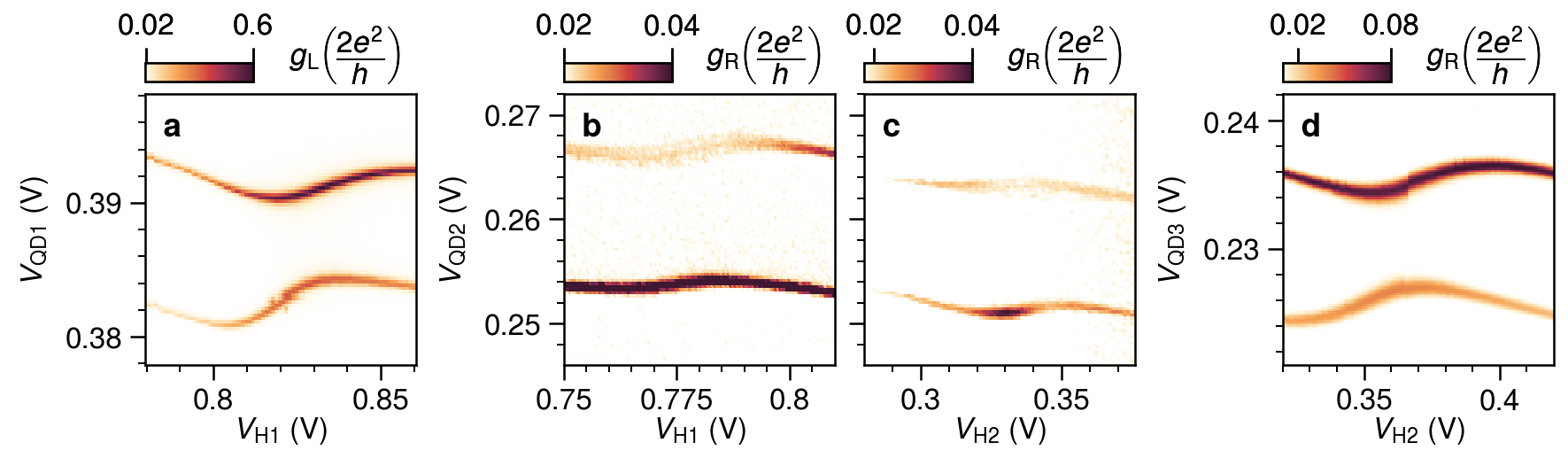}
    \caption{\textbf{QD-ABS charge stability diagrams.} \textbf{a.} Left zero-bias conductance as a function of $\VQDi$ and $\VHi$. \textbf{b, c, d.} Right zero-bias conductance as a function of $\VQDii$ and $\VHi$ (panel c), $\VQDii$ and $\VHii$ (panel c), $\VQDiii$ and $\VHii$ (panel d). All panels show how a pair of QD resonances is modulated by the neighboring hybrid gates, indicating QD-ABS hybridization~\cite{zatelli2023robust}. Panels b and c are measured before forming a QD on the right; here there is a single tunneling barrier separating the right normal lead and the right hybrid so that it is possible to perform spectroscopy of $\QDii$ from the right normal lead as long as the right bias $\VR$ is smaller than the superconducting gap~\cite{florian2023}.}
    \label{supp:s-shapes}
\end{figure}

\begin{figure}[ht!]
    \centering
    \includegraphics[width=\columnwidth]{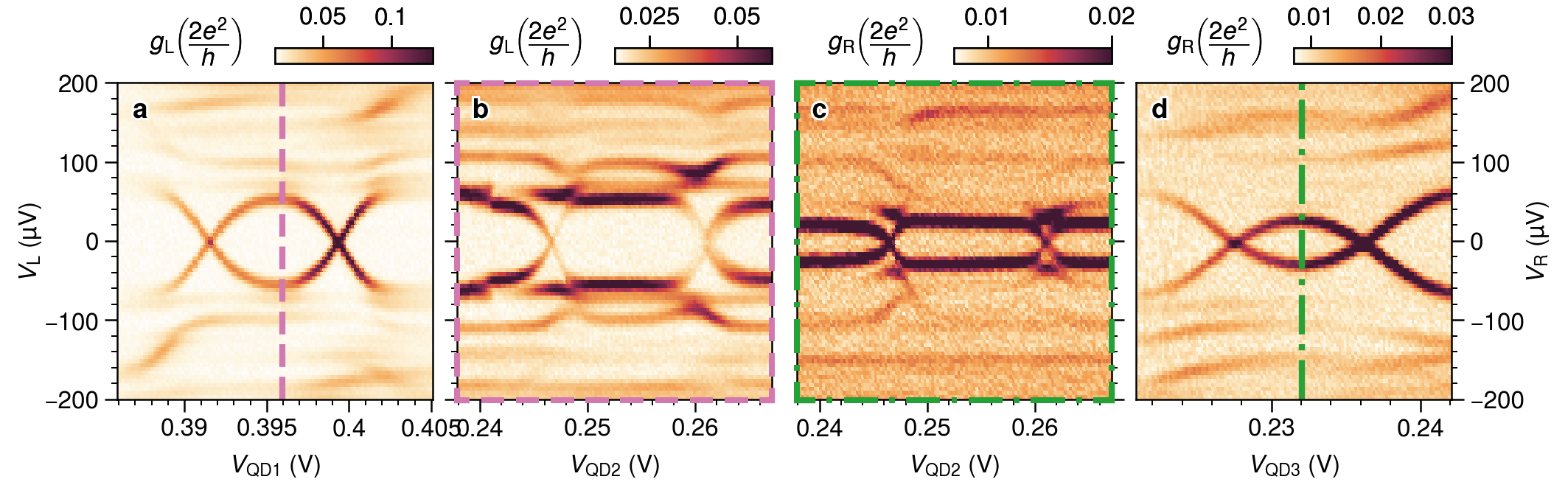}
    \caption{\textbf{QD spectroscopy.} \textbf{a.} $\QDi$ spectroscopy. The QD state appears as an eye-shape, while the ABSs of the left hybrid are visible at higher energies~\cite{zatelli2023robust}. \textbf{b.} $\QDii$ spectroscopy taken from the left probe. Here $\QDi$ is kept in the middle of the pair of charge degeneracy points shown in panel a: $\VQDi = \SI{0.396}{V}$. $\QDi$ states appear as persistent lines at $\approx \pm \SI{60}{\micro V}$ and mix with the ABS and $\QDii$ spectra. \textbf{c.} $\QDii$ spectroscopy taken from the right probe. Here $\VQDiii = \SI{0.232}{V}$. $\QDiii$ states appear as persistent lines at $\approx \pm \SI{25}{\micro V}$ and mix with the ABS and $\QDii$ spectra. Both panels show zero energy crossings at $\approx 0.246$ and $\approx\SI{0.2615}{V}$, which we attribute to $\QDii$ charge transitions. \textbf{d.} $\QDiii$ spectroscopy. We note that the eye-shape is smaller compared to $\QDi$.}
    \label{supp:QD-spectroscopy}
\end{figure}

\begin{figure}[ht!]
    \centering
    \includegraphics[width=0.7\columnwidth]{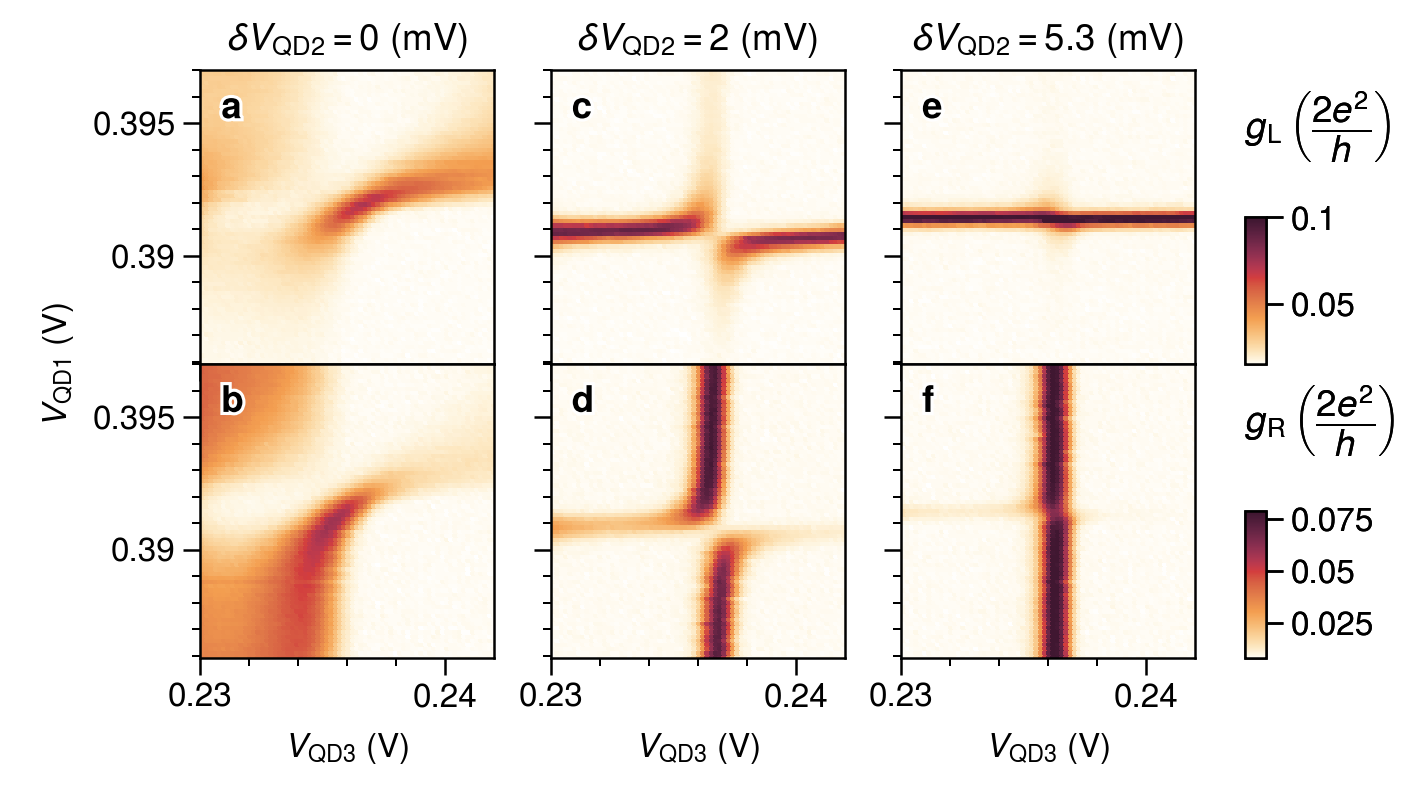}
    \caption{\textbf{Impact of $\QDii$ on the coupling between the outer dots.} \textbf{a-b.} $\QDi$–$\QDiii$ charge stability diagrams measured from the left probe (panel a) and the right one (panel b). At the center of the avoided crossing, all QDs are on resonance. \textbf{c-d.} Same measurements of panels a and b, but with $\QDii$ $\SI{2}{mV}$ off-resonance. \textbf{e-f.} Here $\QDii$ is $\SI{5.3}{mV}$ off-resonance. The more $\QDii$ is detuned, the smaller the $\QDi$–$\QDiii$ avoided crossings are. In panels e and f, the avoided crossings are barely noticeable, indicating suppression of the coupling between the outer dots. Finally, we note that not only the size of the avoided crossings but also the amount of conductance indicates suppression of the $\QDi$–$\QDiii$ coupling: in panel e, the vertical conductance line representing the $\QDiii$ charge transition is barely visible; while in panel f it is the horizontal line corresponding to the $\QDi$ transition to be suppressed.}
    \label{supp:QD2-detuning}
\end{figure}

\begin{figure}[ht!]
    \centering
    \includegraphics[width=0.5\columnwidth]{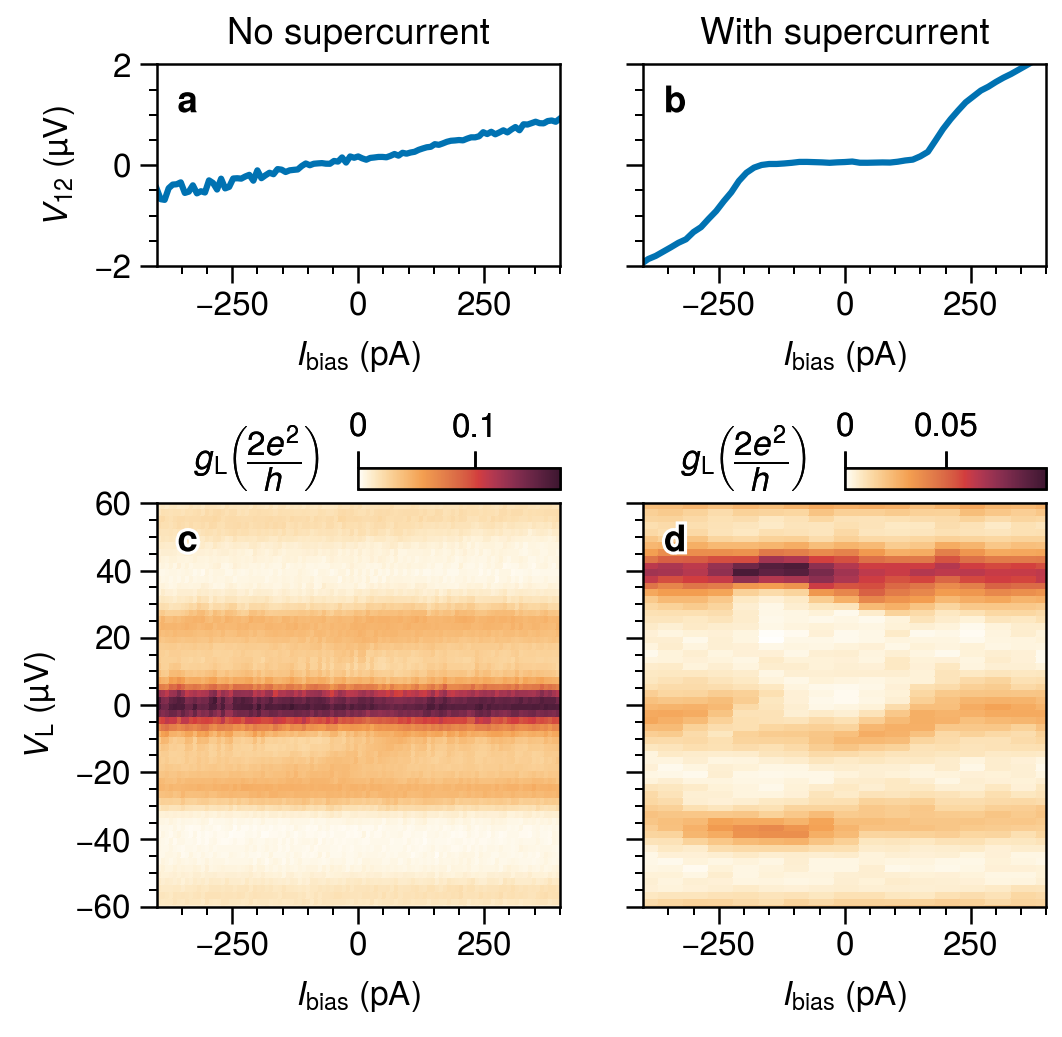}
    \caption{\textbf{a, b.} I-V curves without and with a measurable supercurrent. The voltage $V_{12}$ between the two superconductors is measured as a function of the current bias $\Ibias$ between them. See the code in the shared repository for measurement details. \textbf{c, d.} Left spectroscopy of a three-site chain as a function of $\Ibias$, without and with a measurable supercurrent.
    To avoid complications due to supercurrent, in all the measurements reported in this manuscript (apart from Fig.~\ref{supp:supercurrent}b,d) the tunneling barriers forming $\QDii$ are kept high enough to suppress the supercurrent. In the left column of this figure, we check that such settings show a linear I-V curve and that the $\Ibias$ doesn't affect the three-site chain spectrum.
    To prove that our device can carry supercurrent and this might affect the spectrum of a three-site chain, we lower the $\QDii$ tunneling barriers and measure what is presented in panels b and d. A detailed investigation of the effects of supercurrent is beyond the scope of this manuscript and is left for follow-up works.}
    \label{supp:supercurrent}
\end{figure}

\begin{figure}[ht!]
    \centering
    \includegraphics[width=0.5\columnwidth]{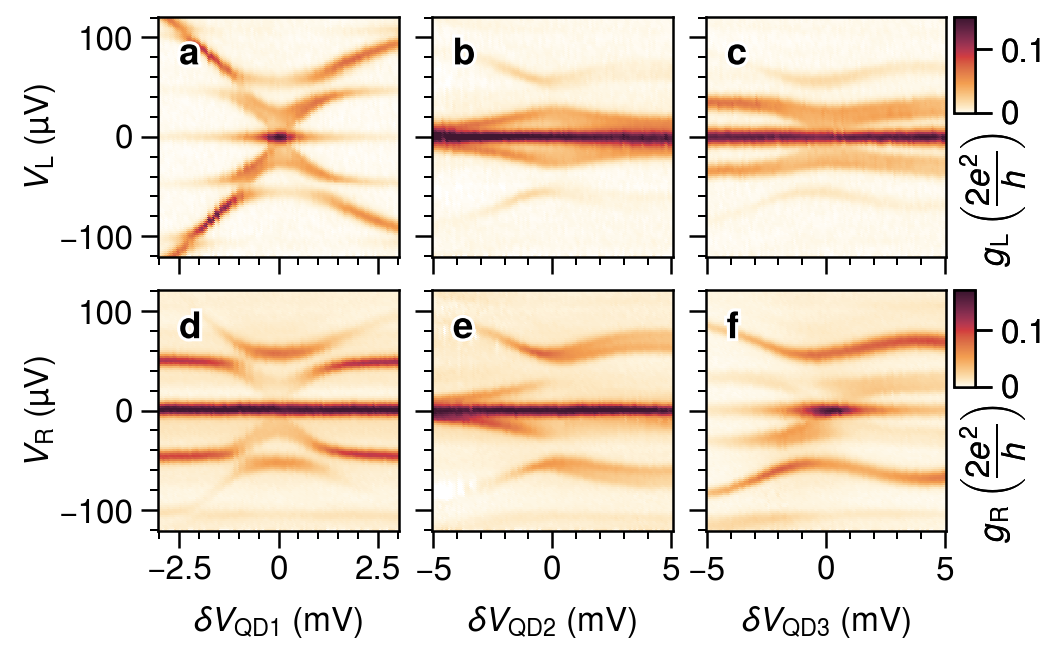}
    \caption{\textbf{Second device. a-f.} Left and right tunneling spectroscopy of a second device tuned to the double sweet-spot condition of Eq.~\ref{eq:double-ss}. Here, $t_1 = \Delta_1 \approx \SI{25}{\micro eV}$ and  $t_2 = \Delta_2 \approx \SI{50}{\micro eV}$. Such coupling strengths are tuned on purpose to values similar to the ones measured for the device in the main text. The remarkable similarity between this figure and panels a-f of Fig.~3 evidences the determinism and reproducibility of our tuning procedure across multiple devices. This device's QDs, ECT and CAR are characterized at zero external magnetic field in ref.~\cite{bordin2024crossed}.}
    \label{supp:second-device}
\end{figure}

\begin{figure}[ht!]
    \centering
    \includegraphics[width=0.66\columnwidth]{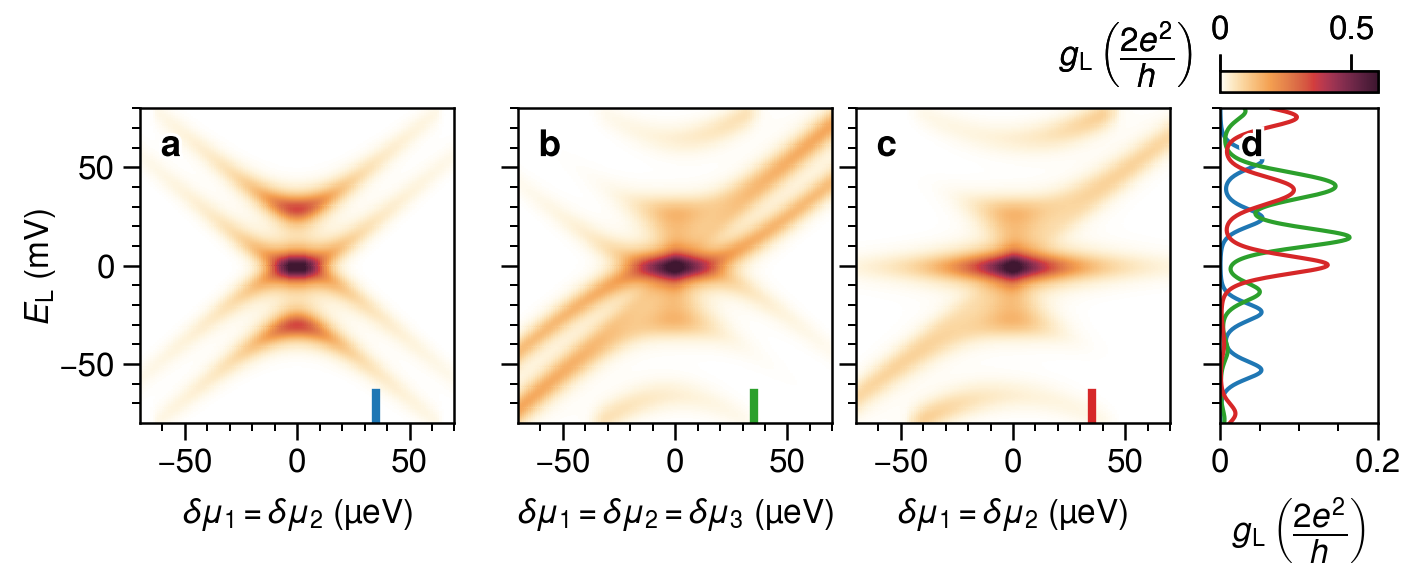}
    \caption{\textbf{a-d.} Theoretical simulation of Fig.~4 data with the spinless model of Eq.~\ref{eq:hamiltonian}. All panels report the average conductance of 25 simulations with different phases on $\Delta_2$. The phase choices are uniformly distributed between 0 and $2\pi$. Panel d displays linecuts at $\mu = \SI{35}{\micro eV}$.}
    \label{supp:protection-mu-theory}
\end{figure}

\begin{figure}[ht!]
    \centering
    \includegraphics[width=\columnwidth]{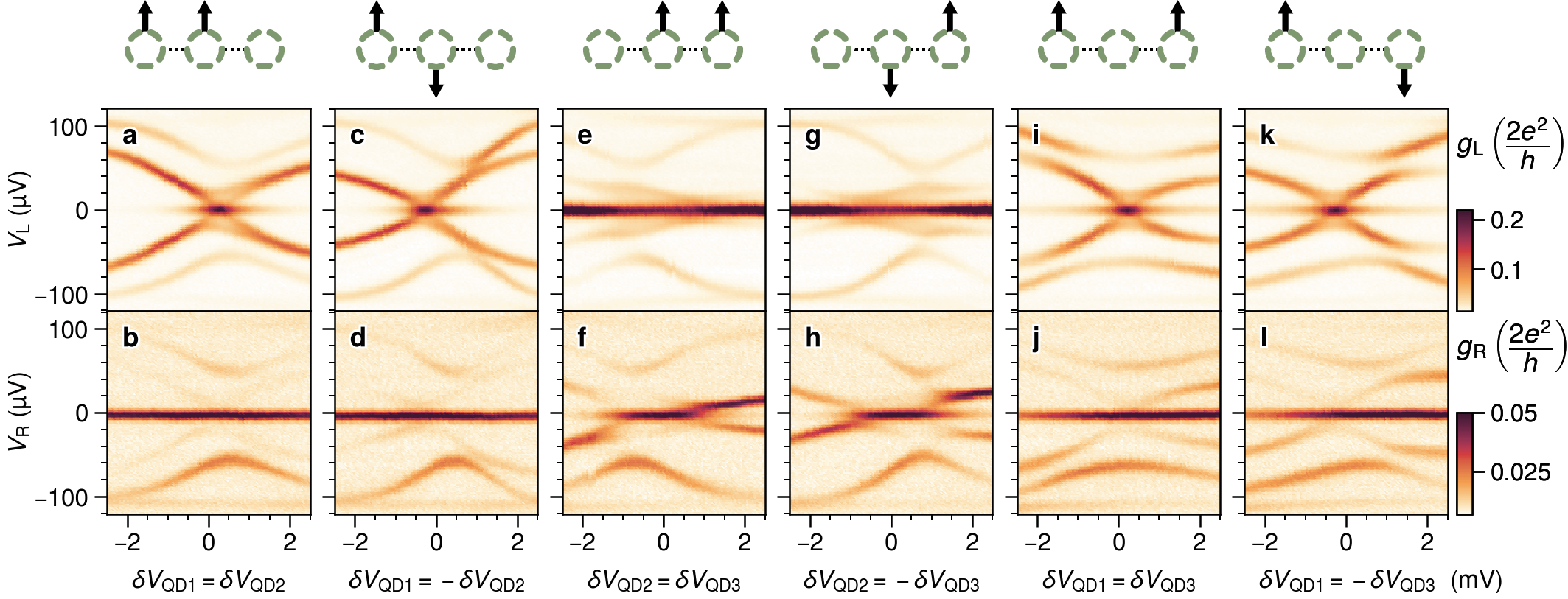}
    \caption{\textbf{Zero bias peak persistence of a three-site chain while detuning any pair of QDs. a-l.} Tunnelling spectroscopy from the left probe (top row) and the right one (bottom row). \textbf{a, b.} Symmetric detuning of $\QDi$ and $\QDii$. \textbf{c, d.} Anti-symmetric detuning of $\QDi$ and $\QDii$. \textbf{e-l.} Symmetric and anti-symmetric detuning of any other pair of QDs. All panels show a persistent zero-bias conductance peak over the full detuning range.}
    \label{supp:bidiags}
\end{figure}

\begin{figure}[ht!]
    \centering
    \includegraphics[width=0.5\columnwidth]{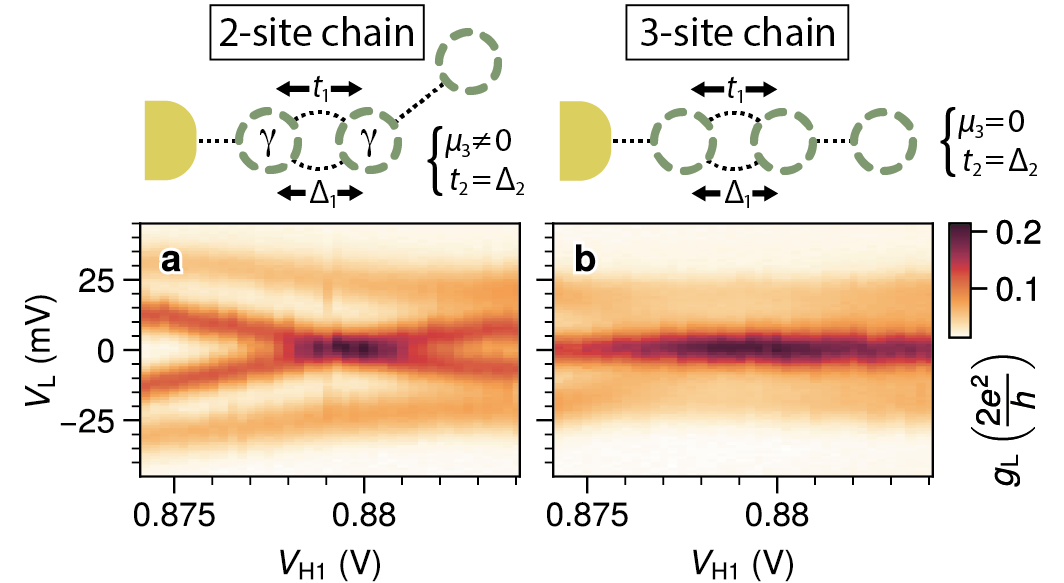}
    \caption{\textbf{a, b.} Tunneling spectroscopy of a 2-site chain (panel a) and 3-site chain (panel b) as a function of $\VHi$. This measurement is a repetition of what is presented in Fig.~\ref{fig:5} of the main text but with higher resolution and only around the $\VHi\approx\SI{0.88}{V}$ sweet-spot. $\delta \VQDiii = -\SI{4}{mV}$ in panel a and $\SI{0}{mV}$ in panel b.}
    \label{supp:t-zoom}
\end{figure}

\begin{figure}[ht!]
    \centering
    \includegraphics[width=0.5\columnwidth]{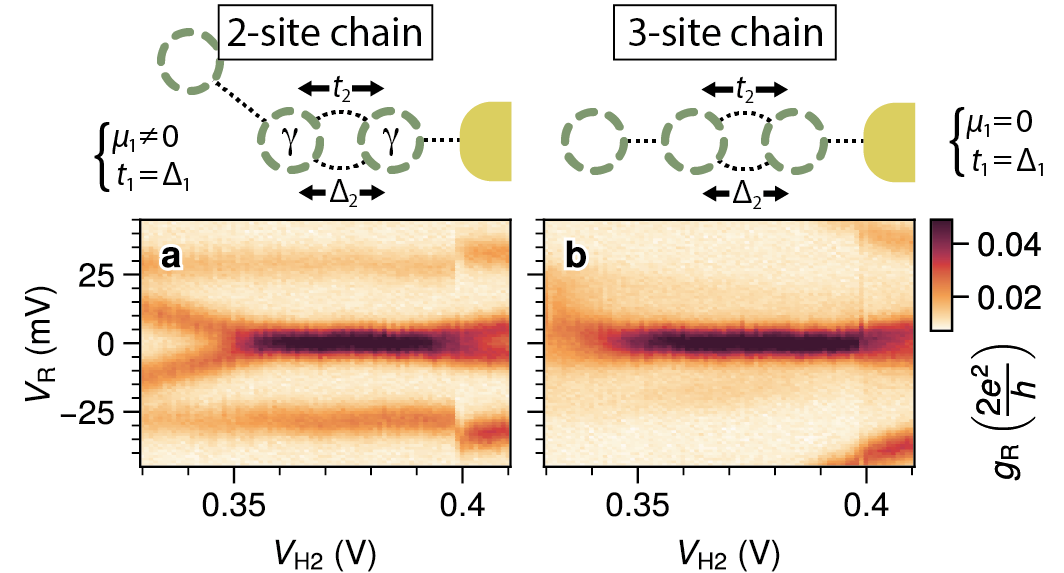}
    \caption{\textbf{a, b.}. Right tunneling spectroscopy of a 2-site chain (panel a) and 3-site chain (panel b) as a function of $\VHii$. $\delta \VQDi = -\SI{4}{mV}$ in panel a and $\SI{0}{mV}$ in panel b. We note that the zero-bias conductance peak of panel a is more stable compared to the one of Fig.~\ref{fig:5}a, we speculate that this is due to accidental similar dispersion of $t_2$ and $\Delta_2$ as a function of $\VHii$. Nevertheless, the 3-site zero-bias conductance peak of panel b is more persistent. We note that such peak broadens and its intensity fades at the edges of the scan, which may indicate the onset of splitting. This is not surprising given the wide $\VHii$ range of this scan ($\SI{80}{mV}$), which can affect the gate configuration due to cross-capacitance.}
    \label{supp:t2}
\end{figure}

\begin{figure}[ht!]
    \centering
    \includegraphics[width=0.5\columnwidth]{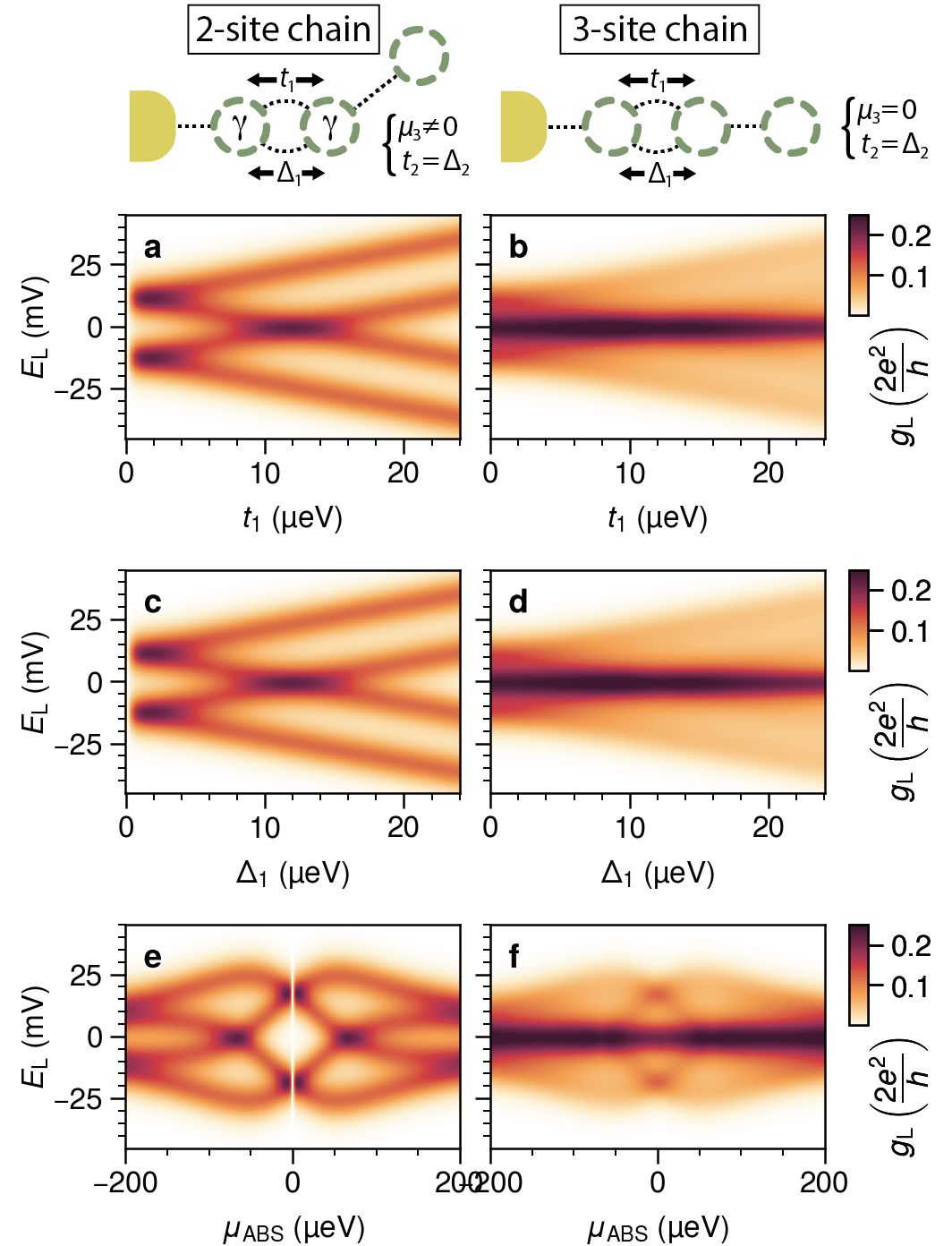}
    \caption{\textbf{a, b.}. Conductance simulation with $t_1$ varied from 0 to \SI{24}{\micro eV} for a 2-site chain (panel a) and a 3-site chain (panel b) exactly as Fig.~5c,d. $\Delta_1 = \SI{12}{\micro eV}$. \textbf{c, d.}. Conductance simulations as a function of $\Delta_1$. $t_1 = \SI{12}{\micro eV}$. The result is identical to the panels above where $t_1$ was varied instead. \textbf{e, f.}. Conductance simulations in a more realistic scenario, where $t_1$ and $\Delta_1$ are varied simultaneously as if there were a single ABS mediating them~\cite{liu2022tunable, bordin2023tunable}.
    In all scenarios, the left column – corresponding to 2-site chains – exhibits zero energy crossings, while the right column – corresponding to 3-site chains – shows persistent zero-bias peaks over the full range. 
    For 3-site chain simulations, the conductance is averaged over 25 phase values of $\Delta_2$, uniformly distributed from 0 to $2\pi$.}
    \label{supp:protection-theory}
\end{figure}

\begin{figure}[ht!]
    \centering
    \includegraphics[width=0.55\columnwidth]{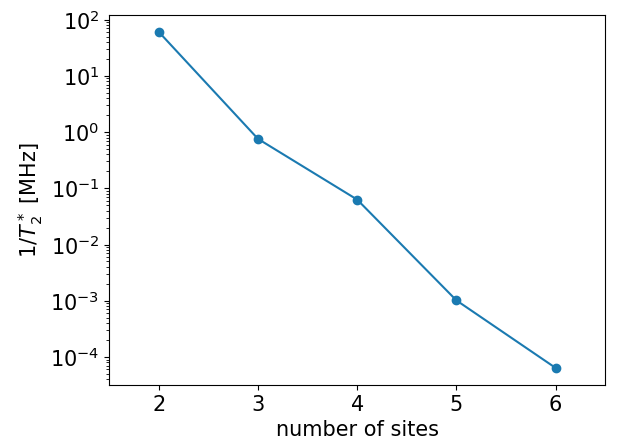}
    \caption{\textbf{Qubit coherence times as a function of the number of sites, assuming charge noise to be the only source of noise.} For fair comparison, we assume homogeneity in the Hamiltonian parameters: $t=\Delta=\SI{20}{\mu eV}$, $\alpha_D=0.04$, $\partial t / \partial V_H= 5\times 10^{-3}$. }
    \label{supp:more-sites}
\end{figure}

%TC:endignore

\end{document}